\documentclass[a4paper,11pt]{article}
\pdfoutput=1 

\usepackage{jheppub} 

\usepackage{hyperref}
\usepackage[subrefformat=parens]{subcaption}
\usepackage{orcidlink}

\title{Search for lepton-flavor-violating $\tau$ decays into a lepton and a vector meson using the full Belle data sample}

\collaboration{The Belle Collaboration}
  \author{N.~Tsuzuki\,\orcidlink{0000-0003-1141-1908},} 
  \author{K.~Inami\,\orcidlink{0000-0003-2765-7072},} 

  \author{I.~Adachi\,\orcidlink{0000-0003-2287-0173},} 
  \author{H.~Aihara\,\orcidlink{0000-0002-1907-5964},} 
  \author{D.~M.~Asner\,\orcidlink{0000-0002-1586-5790},} 
  \author{H.~Atmacan\,\orcidlink{0000-0003-2435-501X},} 
  \author{T.~Aushev\,\orcidlink{0000-0002-6347-7055},} 
  \author{R.~Ayad\,\orcidlink{0000-0003-3466-9290},} 
  \author{V.~Babu\,\orcidlink{0000-0003-0419-6912},} 
  \author{Sw.~Banerjee\,\orcidlink{0000-0001-8852-2409},} 
% ...

\abstract{
Charged-lepton-flavor-violation is predicted in several new physics scenarios.
We update the analysis of $\tau$ lepton decays into a light charged lepton ($\ell$ = $e^{\pm}$ or $\mu^{\pm}$)
and a vector meson ($V^0$ = $\rho^0$, $\phi$, $\omega$, $K^{\ast0}$, or $\overline{K}{}^{\ast0}$)
using 980 fb$^{-1}$ of data collected with the Belle detector at the KEKB collider.
No significant excess of such signal events is observed, and thus 90\% credibility level upper limits are set
on the $\tau \rightarrow \ell V^0$ branching fractions in the range of (1.7--$4.3) \times 10^{-8}$.
These limits are improved by 30\% on average from the previous results.
}

\arxivnumber{2301.03768}

\begin{document}
\maketitle
\flushbottom

\section{Introduction}
\label{sec:intro}
In the Standard Model, charged-lepton-flavor-violation (CLFV) is so strongly suppressed that it is undiscoverable by current experiments.
Therefore, a discovery of a CLFV event indicates new physics (NP).
Verifying various NP models requires many searches of various CLFV modes~\cite{Celis:2014asa}.
Whereas the CLFV constraints are much more stringent for $\mu$-to-$e$ than for $\tau$ through the precise measurements~\cite{MEG:2016leq,SINDRUM:1987nra,SINDRUMII:2006dvw},
we are interested in $\tau$, the third-generation and heaviest lepton.
So-called $B$-anomalies, which indicate NP effects in $B$ semileptonic decays~\cite{BaBar:2012obs,BaBar:2013mob,Belle:2015qfa,Belle:2016dyj,Belle:2019rba,LHCb:2015gmp,LHCb:2017smo,LHCb:2017rln,Belle:2016fev,LHCb:2020lmf,LHCb:2020gog,LHCb:2021zwz}, also motivate the CLFV searches.

We focus on $\tau$ CLFV decays into a charged lepton ($\ell$ = $e^{\pm}$ or $\mu^{\pm}$) and a neutral vector meson ($V^0$ = $\rho^0$, $\phi$, $\omega$, $K^{\ast0}$, or $\overline{K}{}^{\ast0}$).
In refs.~\cite{Hati:2020cyn,DiLuzio:2018zxy,Kumar:2018kmr,Crivellin:2018yvo,Crivellin:2019dwb,BhupalDev:2020zcy}, the $\tau \rightarrow \mu \phi$ mode is a sensitive probe for leptoquark models that can explain the $B$-anomalies.\footnote{
One of the $B$-anomalies which motivated the models described in those references is the $R(K^{(\ast)})$ anomaly reported by the LHCb experiment~\cite{LHCb:2021trn}, but it disappeared in their updated analysis~\cite{LHCb:2022qnv}.}
Some other NP models predict branching fractions of $\mathcal{O}(10^{-10})$--$\mathcal{O}(10^{-8})$ for $\tau \rightarrow \ell V^0$~\cite{Ilakovac:1999md,Li:2009yr,Arhrib:2009xf,Pacheco:2022ebc}.

We previously searched for $\tau \rightarrow \ell V^0$ events using 854 fb$^{-1}$ of Belle data, and set 90\% credibility level (C.L.) upper limits on the branching fractions in the range of $(1.2$--$8.4)\times 10^{-8}$~\cite{Belle:2011ogy}.\footnote{
In common high energy physics usage, this credibility level has been reported as ``confidence level,'' which is a frequentist-statistics term.}
This paper reports an updated search for $\tau \rightarrow \ell V^0$ using the full 980 fb$^{-1}$ Belle data set.
The signal efficiency is improved through new event selection criteria,
which allow more signal candidates first and then remove background events by a multivariate analysis.

\section{Belle experiment}
\label{sec:belle}
The Belle detector is a spectrometer that covers large solid angles of the $e^+e^-$ collision events from the KEKB accelerator~\cite{Kurokawa:2001nw,Abe:2013kxa}.
The detector consists of a silicon vertex detector, a 50-layer central drift chamber (CDC), an array of aerogel threshold Cherenkov counters, time-of-flight scintillation counters, and an electromagnetic calorimeter composed of 8736 CsI(Tl) crystals (ECL).
These devices are located inside a superconducting solenoid coil that provides a 1.5 T magnetic field.
An iron flux return located outside of the coil is instrumented to detect $K_L^0$ mesons and identify muons.
The Belle detector is described in detail elsewhere~\cite{Belle:2000cnh,Belle:2012iwr}.

Of the 980 fb$^{-1}$ data set, 703 fb$^{-1}$ was collected at the $\Upsilon$(4S) resonance, 121 fb$^{-1}$ at the $\Upsilon$(5S), 89 fb$^{-1}$ at an energy 60 MeV below the $\Upsilon$(4S), 28 fb$^{-1}$ of energy-scans above the $\Upsilon$(4S), and the remainder at and near the $\Upsilon$(1--3S).
Compared to the previous paper~\cite{Belle:2011ogy}, the following data sets have been added: 78 fb$^{-1}$ at and near the $\Upsilon$(5S), 38 fb$^{-1}$ at and near the $\Upsilon$(1--3S), and 10 fb$^{-1}$ at an energy 60 MeV below the $\Upsilon$(4S).

The $e^+e^-$ collision events in the Belle detector are simulated by the Monte Carlo (MC) method.
Signal MC events of $\tau \rightarrow \ell V^0$ are generated by a dedicated MC with KKMC and TAUOLA~\cite{Jadach:1999vf}, where $\tau^+\tau^-$ pairs are initially produced and one of the $\tau$'s decays into $\ell V^0$ and the other decays generically.
The numbers of generated signal MC events are $1.1\times 10^{6}$ events at the $\Upsilon$(4S) resonance, $0.4\times 10^{6}$ events at the $\Upsilon$(5S), $0.1\times 10^{6}$ events at each of the $\Upsilon$(1--3S), and $0.1\times 10^{6}$ events at an energy 60 MeV below the $\Upsilon$(4S).
We assume a uniform CLFV decay angle in the $\tau$ rest frame.
No specific NP model is assumed in the CLFV decay process, and the spin direction of $V^0$ is set randomly and independently of the spin of the mother $\tau$.
For background MC simulations, $e^+e^- \rightarrow q \bar{q}$ ($q=u, d, s, c$), $e^+e^- \rightarrow \tau^+ \tau^-$, Bhabha, and two-photon processes are generated by EvtGen~\cite{Lange:2001uf}, KKMC~\cite{Jadach:1999vf}, BHLUMI~\cite{Jadach:1991by}, and AAFH~\cite{Berends:1986ig}, respectively.
The detector responses are simulated by GEANT3~\cite{Brun:1082634}.

\section{Reconstruction and event selection}
\label{sec:selection}
To improve the signal selection efficiency, we loosen or change the event selection criteria in the previous paper~\cite{Belle:2011ogy} for this analysis.
Some of the loose selection variables are used as inputs to the multivariate analysis, which is intended to reduce background events.
All the event selection criteria have been optimized and are described in this section.

A signal $\tau$ is reconstructed from a lepton and a neutral vector meson.
We separate the event into two hemispheres in the center-of-mass (c.m.) frame by a plane perpendicular to the thrust vector ($\vec{n}_{T}$)~\cite{Brandt:1964sa,Farhi:1977sg}.
The thrust vector is obtained by maximizing the thrust $T=\Sigma_{i}|\vec{p}_{i}^{\ \mathrm{c.m.}} \cdot \vec{n}_{T}| / \Sigma_{i}|\vec{p}_{i}^{\ \mathrm{c.m.}}|$, where $i$ runs over all tracks and photons, and $\vec{p}_{i}^{\ \mathrm{c.m.}}$ is the momentum in the c.m. frame.
Here, photons with energies above 0.1 GeV are used.
In the hemisphere that contains a $\tau$ CLFV decay (called signal side and $\tau_{\mathrm{sig}}$), $V^0$ is reconstructed as follows:
$\rho^0$ from $\pi^+\pi^-$ within the reconstructed mass window of 0.445--1.08 GeV/$c^2$, $\phi$ from $K^+K^-$ within 1.00--1.04 GeV/$c^2$, $\omega$ from $\pi^+\pi^-\pi^0$ within 0.7--0.9 GeV/$c^2$, $K^{\ast0}$ from $K^+\pi^-$ within 0.7--1.1 GeV/$c^2$, and $\overline{K}{}^{\ast0}$ from $K^-\pi^+$ within 0.7--1.1 GeV/$c^2$.
In the other hemisphere (called tag side), the other $\tau$ ($\tau_{\mathrm{tag}}$) is reconstructed from $\ell^{\pm}\nu\nu$, $\pi^{\pm}\nu$, $\pi^{\pm}\pi^0\nu$, $\pi^{\pm}\pi^0\pi^0\nu$, or $\pi^{\pm}\pi^{\mp}\pi^{\pm}\nu$.
The $\tau_{\mathrm{tag}} \rightarrow \pi^{\pm}\pi^0\pi^0\nu$ and $\pi^{\pm}\pi^{\mp}\pi^{\pm}\nu$ decays were discarded in the previous paper~\cite{Belle:2011ogy}.
This $\tau_{\mathrm{tag}}$ information enables the suppression of background events that have no neutrinos in the tag side.

The signal $\tau \rightarrow \ell V^0$ events have a unique kinematical feature;
the $\ell V^0$ invariant mass ($M_{\ell V^0}$) is close to the $\tau$ mass and the difference of the $\ell V^0$ energy from the beam energy in the c.m. frame ($\Delta E$) is close to zero.
The signal events within 1.65 GeV/$c^2$ $< M_{\ell V^0} <$ 1.90 GeV/$c^2$ and $|\Delta E| <$ 0.5 GeV are reconstructed in this paper.
We follow a blind analysis approach in this search by not looking at the signal candidates in the data set until finalizing the event selection and background estimation.
The blind region is 1.75 GeV/$c^2$ $\leq M_{\ell V^0} <$ 1.81 GeV/$c^2$ and $|\Delta E| <$ 0.08 GeV for the $\mu \rho^0$, $\mu \phi$ and $\mu K^{\ast0}$($\overline{K}{}^{\ast0}$) modes,
and 1.74 GeV/$c^2$ $\leq M_{\ell V^0} <$ 1.82 GeV/$c^2$ and $|\Delta E| <$ 0.1 GeV for the other modes.

Charged tracks, photons, and $\pi^0$s should satisfy the following selection criteria.
Each charged track or photon is within the fiducial volume defined by $-0.866 < \cos\theta < 0.956$,
where $\theta$ is the polar angle with respect to the direction opposite to the $e^+$ beam in the laboratory frame.
Charged tracks come from the interaction point;
the distance of the closest point from the interaction point is less than 0.5 cm in the transverse direction and less than 3.0 cm in the longitudinal direction.
Each $\pi^0$ is reconstructed from two photons inside the same hemisphere and the photon energy ($E_{\gamma}$) should be larger than 0.05 GeV.
The $\pi^0$ mass window is 0.12 GeV/$c^2$ $< M_{\gamma\gamma} <$ 0.15 GeV/$c^2$, corresponding to $\pm3\sigma$ in the $\pi^0$ mass resolution.
A $\pi^0$ mass-constrained fit is performed to improve the energy resolution.

After reconstructing the signal and tag $\tau$'s, no extra charged tracks are allowed.
We count the number of photons ($n_{\gamma}$) with $E_{\gamma}$ larger than 0.1 GeV in the signal side, and require $n_{\gamma} \leq$ 3 for the $\ell\omega$ mode, which includes a $\pi^0 \rightarrow \gamma\gamma$, and $n_{\gamma} \leq$ 1 for the other modes.

Particle identification is effective in suppressing the main background events of three-hadron-track to the $\tau \rightarrow \ell V^0$ signal.
We use likelihood ratios for electron identification ($\mathcal{P}(e)$)~\cite{Hanagaki:2001fz} and muon identification ($\mathcal{P}(\mu)$)~\cite{Abashian:2002bd}.
The lepton identification criteria are $\mathcal{P}(e) >$ 0.9 for electrons, and $\mathcal{P}(\mu) >$ 0.95 and the momentum is larger than 0.6 GeV/$c$ for muons.
The electron (muon) identification efficiency is 90\% (75\%), whereas the probability of misidentifying a pion as an electron (muon) is 0.1\% (2\%).
These efficiencies and fake rates are different from the previous paper~\cite{Belle:2011ogy}, because the track momentum requirements are loosened in this paper.
The energy loss of an electron by bremsstrahlung is recovered by adding back the energy of every photon within 0.05 radians from the electron track direction into the electron momentum.
To suppress low-multiplicity background events like Bhabha, $ee \rightarrow eeee$, or $ee \rightarrow ee\mu\mu$,
an electron veto ($\mathcal{P}(e) <$ 0.9) is applied to all hadron candidate tracks from $V^0$ and $\tau_{\mathrm{tag}}$.

For hadron identification, we use a binary likelihood ratio $\mathcal{P}(i|j)=\mathcal{L}_i/(\mathcal{L}_i+\mathcal{L}_j)$, where $\mathcal{L}_{i (j)}$ is the likelihood of particle $i$ ($j$)~\cite{Nakano:2002jw} and $i$ ($j$) is $\pi,\ K$, or $p$.
The kaon identification criteria are $\mathcal{P}(K|\pi) >$ 0.6 for both charged kaons from $\phi$ decay and $\mathcal{P}(K|\pi) >$ 0.8 for the charged kaon from $K^{\ast0}$ and $\overline{K}{}^{\ast0}$ decays.
The kaon identification efficiency is 86\% (77\%), whereas the probability of misidentifying a charged pion as a kaon is 4\% (2\%) for the kaons from $\phi$ ($K^{\ast0}$, $\overline{K}{}^{\ast0}$).
A kaon veto ($\mathcal{P}(K|\pi) <$ 0.6) is applied to both charged pions from $\rho^0$ in the signal side, and 96\% of pions are retained, whereas 14\% of kaons are not vetoed.
We do not apply this kaon veto for the charged pions from $\omega$, $K^{\ast0}$, and $\overline{K}{}^{\ast0}$.
To suppress muons from kaons decaying inside the CDC ($K^{\pm} \rightarrow \mu^{\pm}\nu$), 
the kaon veto is also applied to the signal-side muon track for the hadronic tags ($\tau_{\mathrm{tag}}^{\pm} \rightarrow \pi^{\pm}\nu,\ \pi^{\pm}\pi^0\nu,\ \pi^{\pm}\pi^{\mp}\pi^{\pm}\nu$, or $\pi^{\pm}\pi^0\pi^0\nu$).
For the $\mu V^0$ modes with the hadronic tags, a proton veto ($\mathcal{P}(p|K) <$ 0.6 and $\mathcal{P}(p|\pi) <$ 0.6) is applied for the tag-side tracks.

The signal events have one or two neutrinos from the $\tau_{\mathrm{tag}}$ decay.
We introduce some event selection criteria requiring one or more neutrinos in the tag side.
The missing momentum due to the neutrino(s) is calculated by subtracting the vector sum of the momenta of all tracks and photons from the sum of the beam momenta.
The missing energy is also calculated by subtracting the sum of the energy of all tracks and photons from the sum of the beam energy.
Here, extra photons that are not used for the $\tau$ reconstruction are included.
The transverse missing momentum is required to be larger than 0.5 GeV/$c$, and the missing energy in the c.m. frame ($E_{\mathrm{miss}}^{\mathrm{c.m.}}$) is required to be larger than 0 GeV.
Events with missing particles other than neutrinos should be rejected as background events.
These non-neutrino missing particles can arise in two ways: neutral particles pass through the gaps between the barrel and end-cap ECLs, and any particles go outside the CDC volume.
Thus, the direction of the missing momentum is required not to point to such regions.
The missing particles should be in the tag side and hence $\cos\theta_{\mathrm{miss-tag}}^{\mathrm{c.m.}} >$ 0, where $\theta_{\mathrm{miss-tag}}^{\mathrm{c.m.}}$ is the angle between the missing momentum and the vector sum of the momenta of the tag-side tracks and photons in the c.m. frame.
The neutrino angle with respect to the $\tau_{\mathrm{tag}}$ momentum direction is restricted in a $\tau_{\mathrm{tag}}$ two-body decay;
thus $\cos\theta_{\mathrm{miss-tag}}^{\mathrm{c.m.}} <$ 0.85 is also applied for the $\ell \rho^0$ modes with $\tau_{\mathrm{tag}}^{\pm} \rightarrow \pi^{\pm} \nu$.

We require features of a generic $\tau$ decay in the tag side.
The invariant mass of the particles including all photons in the tag hemisphere should be less than the $\tau$ mass (1.777 GeV/$c^2$).
For $\tau_{\mathrm{tag}}$ decays into $\pi^{\pm}\pi^0\nu$ ($\pi^{\pm}\pi^{\mp}\pi^{\pm}\nu$ or $\pi^{\pm}\pi^0\pi^0\nu$), the reconstructed mass of those pions is required to be 0.4 GeV/$c^2$ $< M_{\pi^{\pm}\pi^0} <$ 1.3 GeV/$c^2$ (0.7 GeV/$c^2$ $< M_{3\pi} <$ 1.7 GeV/$c^2$), which corresponds to the mass of $\rho^{\pm}$ ($a_1^{\pm}$).

After the above event reconstruction, the background sources are the $q\bar{q}$ continuum ($q=u,d,s,c$), generic $\tau^+ \tau^-$, and low-multiplicity events.
The low-multiplicity events especially contribute to the background events for $eV^0$ modes that have electron tracks.
We suppress the low-multiplicity events first, and then use a maltivariate analysis tool to suppress the $q\bar{q}$ continuum and generic $\tau^+ \tau^-$ events.

The Bhabha events have tracks from photon conversion.
To suppress these background events for the $eV^0$ modes, the invariant mass of the electron and one of the tracks from the $V^0$, assigned the electron-mass hypothesis, should be larger than 0.2 GeV/$c^2$.
In addition, for the $eK^{\ast0}$ and $e\overline{K}{}^{\ast0}$ modes, the invariant mass of the two tracks from the $V^0$, each assigned the electron-mass hypothesis, is required to be larger than 0.1 GeV/$c^2$.
This event selection also suppresses some of the generic $\tau^+\tau^-$ events, which have tracks from photon conversion.

\begin{figure}[htb]
  \centering
  \includegraphics[width=7.5cm,clip]{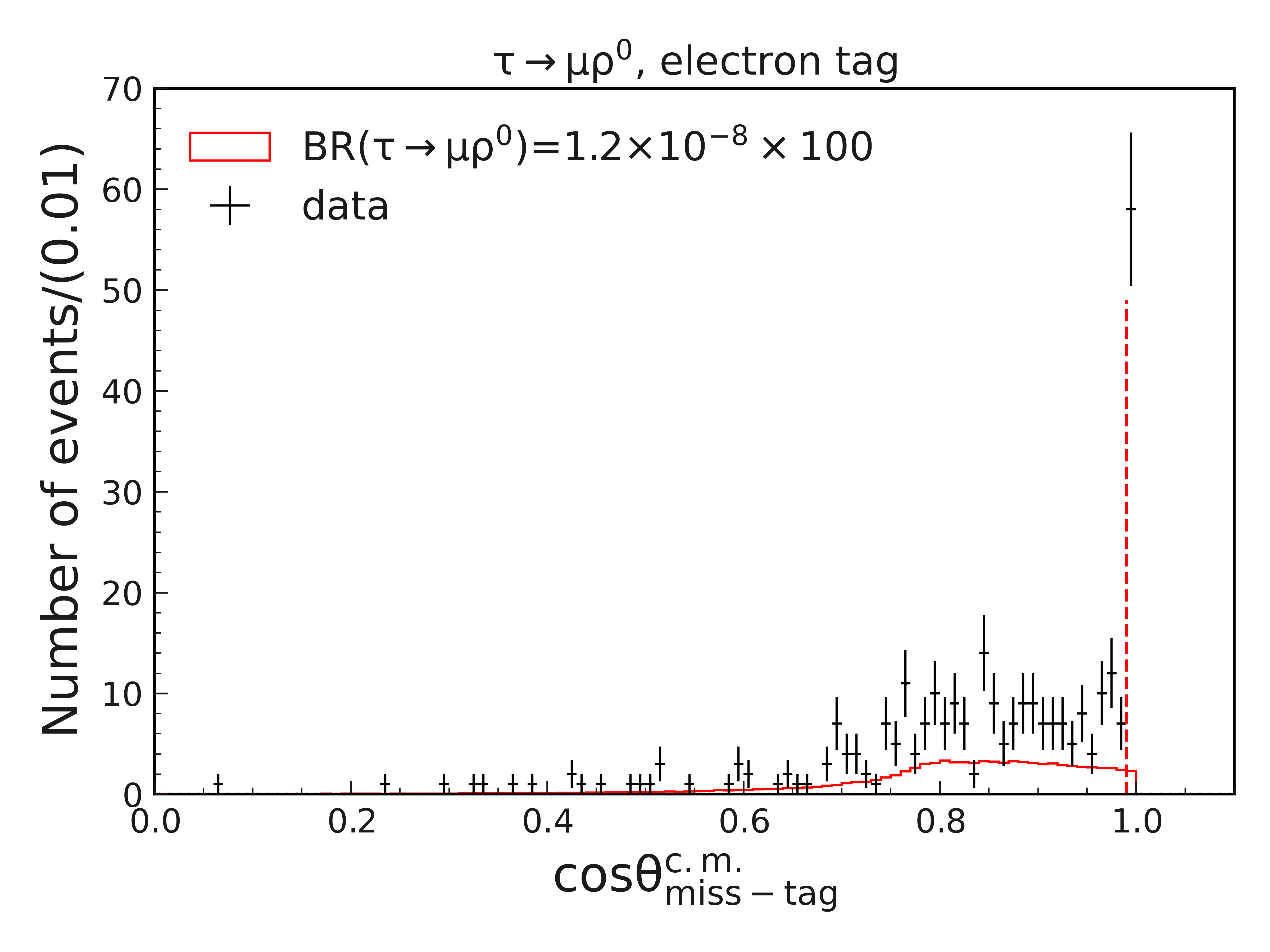}
  \caption{The $\cos\theta_{\mathrm{miss-tag}}^{\mathrm{c.m.}}$ distribution of the $\tau \rightarrow \mu \rho^0$ mode with a electron tag track after the reconstruction, particle identification, and photon conversion event suppression. Black points with error bars are the data outside the blind region. Red solid histogram is the signal MC. The signal MC is scaled to the number of events corresponding to 100 times as large branching fraction as the current upper limit. The red dashed line is the upper limit to remove the low-multiplicity events. The low-multiplicity events cluster around $\cos\theta_{\mathrm{miss-tag}}^{\mathrm{c.m.}} =$ 1, whereas the other background events are linearly distributed in the region of $\cos\theta_{\mathrm{miss-tag}}^{\mathrm{c.m.}} >$ 0.8.}
  \label{fig:cosTheta-mrho}
\end{figure}

The low-multiplicity background events are still not negligible for the events with electrons: $\tau \rightarrow eV^0$ or $\tau_{\mathrm{tag}} \rightarrow e\nu\nu$.
Because the missing particles of the low-multiplicity background events are the bremsstrahlung photons from the electron in the tag side, $\cos\theta_{\mathrm{miss-tag}}^{\mathrm{c.m.}}$ is close to one (Figure~\ref{fig:cosTheta-mrho}).
In addition, the missing energy is small for some low-multiplicity background events.
For the $\mu \rho^0$ mode with $\tau_{\mathrm{tag}}^{\pm} \rightarrow e^{\pm}\nu\nu$, $\cos\theta_{\mathrm{miss-tag}}^{\mathrm{c.m.}} <$ 0.99 and $E_{\mathrm{miss}}^{\mathrm{c.m.}} >$ 0.4 GeV selection criteria are applied.
For the $eV^0$ modes with $\tau_{\mathrm{tag}}^{\pm} \rightarrow e^{\pm}\nu\nu$ or $\pi^{\pm}\nu$, $\cos\theta_{\mathrm{miss-tag}}^{\mathrm{c.m.}} <$ 0.97 is applied.
For the $eV^0$ modes with $\tau_{\mathrm{tag}}^{\pm} \rightarrow e^{\pm}\nu\nu$, $E_{\mathrm{miss}}^{\mathrm{c.m.}}$ should be larger than 0.4, 2.0, and 1.5 GeV for $e\phi$, $e\rho^0$, and the other $eV^0$ modes, respectively.

\begin{figure}[htb]
  \centering
  \includegraphics[width=7.5cm,clip]{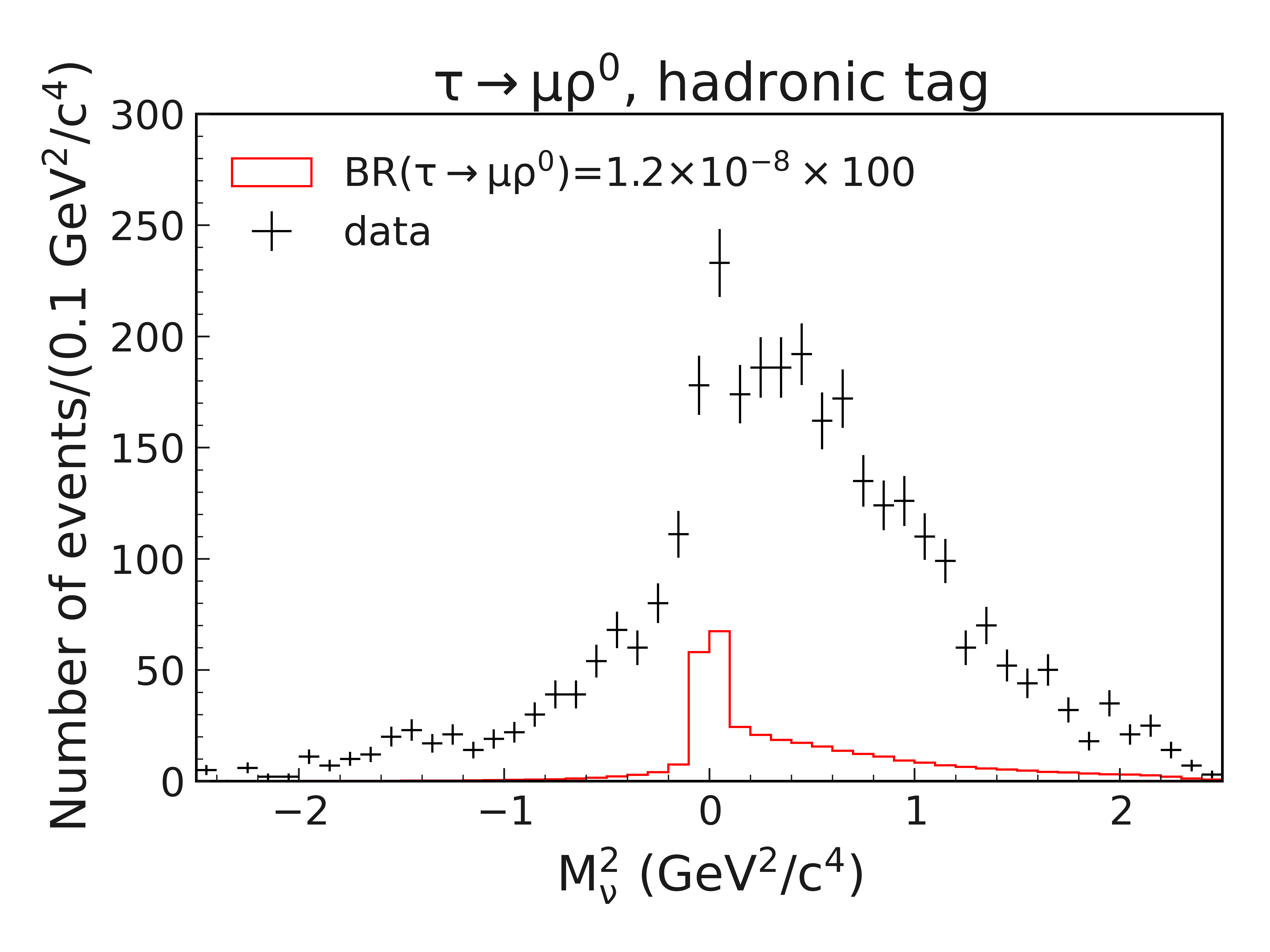}
  \caption{The $M_{\nu}^2$ distribution of the $\tau \rightarrow \mu \rho^0$ mode with the hadronic tags after the event selection except for the requirement of the BDT output. Black points with error bars are the data outside the blind region. Red solid histogram is the signal MC. The signal MC is scaled to the number of events corresponding to 100 times as large a branching fraction as the current upper limit. The events constituting the upper tail of the signal distribution originate from wrong or missing $\pi^0$ in the tag side.}
  \label{fig:M2_miss-mrho}
\end{figure}

The remaining background events are mainly from the $q\bar{q}$ continuum ($q=u,d,s,c$) and generic $\tau^+ \tau^-$ events, which have three charged pion tracks in the signal side.
We use two-class Boosted Decision Tree (BDT) algorithms for signal and these background classification.
The BDT library is LightGBM~\cite{NIPS2017_6449f44a}.
One BDT for each $\ell V^0$ mode is trained.
These BDTs output signal probabilities using the following input variables:
\begin{itemize}
\item $M_{V^0}$, $M_{\nu}^2$, $P_{\nu}^{\mathrm{c.m.}}$, $T$, $P_{\ell}^{\mathrm{sig}}$, $E_{\mathrm{tag}}^{\mathrm{hemi}}$, $\cos\theta_{\mathrm{miss-tag}}^{\mathrm{c.m.}}$
\item (categorical variables) $\tau_{\mathrm{tag}}$ decay mode, collision energy
\item (additional for the $\ell\omega$ modes) $P_{\pi^0}^{\mathrm{sig}}$, $E_{\gamma}^{\mathrm{low}}$, 
\end{itemize}
where $M_{V^0}$ is the invariant mass of the vector meson, $M_{\nu}^2$ is the missing mass squared, $P_{\nu}^{\mathrm{c.m.}}$ is the missing momentum in the c.m. frame, $T$ is the magnitude of the thrust vector~\cite{Brandt:1964sa,Farhi:1977sg}, $P_{\ell}^{\mathrm{sig}}$ is the momentum of the lepton in the signal side, $E_{\mathrm{tag}}^{\mathrm{hemi}}$ is the energy sum of the tracks and photons in the tag hemisphere, $P_{\pi^0}^{\mathrm{sig}}$ is the momentum of $\pi^0$ from $\omega$ and $E_{\gamma}^{\mathrm{low}}$ is the lower energy of the two photons from the $\pi^0$.
The variables of neutrino kinematics ($M_{\nu}^2$ and $P_{\nu}^{\mathrm{c.m.}}$) were not used for the event selection in the previous paper~\cite{Belle:2011ogy}.
They are calculated from the momenta of the reconstructed $\tau_{\mathrm{sig}}$ and $\tau_{\mathrm{tag}}$, where the energy of $\tau_{\mathrm{sig}}$ is fixed to the half of the beam energy in the c.m. frame.
The $q\bar{q}$ continuum background events can be effectively suppressed by a $M_{\nu}^2$ selection in the hadronic tags, involving only one neutrino (Figure \ref{fig:M2_miss-mrho}).

The training, validation and evaluation of the BDTs are done with 40\%, 10\%, and 50\% of the signal MC, respectively.
Regarding the training and validation samples for the background events, we utilize hadron background enhanced data that are obtained by removing the lepton identification for the signal-side leptons but with a lepton identification veto ($\mathcal{P}(e) \leq$ 0.9 and $\mathcal{P}(\mu) \leq$ 0.95) for all the signal-side tracks in the data.
The hadron background enhanced data have a much larger number of events than the background data with the nominal selection criteria, whereas both data sets are composed mainly of three charged pions from $\tau$ decays or from continuum events.
The training is done with 80\% of the hadron background enhanced data and the validation is done with 20\%.
During BDT training, a weight is applied to each of the signal MC events such that the sum of the weights is equal to the number of background events.
We monitor the area under curve (AUC) of the Receiver Operating Characteristic curve~\cite{doi:10.1148/radiology.143.1.7063747} using the validation samples after each training step, and choose the number of training steps with the best AUC score.

The event selection with the BDT output (BDT selection) is determined only by a target signal efficiency.
The target signal efficiency is determined based on the signal efficiency with a cut-based event selection.
In the cut-based event selection, the $M_{V^0}$ windows correspond to $\pm 2\sigma$ of reconstructed mass distribution, and the $M_{\nu}^2$ windows are set for each $\ell V^0$ mode and each $\tau_{\mathrm{tag}}$ decay mode so that the expected number of background events inside the signal region ($N_{\mathrm{BG}}$, see the next section) is approximately one or less.
The target signal efficiency with the BDT selection is set as relatively 5\% larger than that with the cut-based event selection, because we expect improvement in separating the signal events from the background events.

The finalized BDT selection shows similar $N_{\mathrm{BG}}$ to that of the cut-based event selection.
The BDT selection is not applied to the $\ell \phi$ modes because $N_{\mathrm{BG}}$ in each of the two modes is small enough.

\section{Signal efficiency and background estimation}
\label{sec:eff-nbg}
We define the signal region with an ellipse in the $M_{\ell V^0}$--$\Delta E$ plane inside the blind region.
Most of the signal events cluster around $M_{\ell V^0} = 1.777$ GeV/$c^2$ and $\Delta E = 0$ GeV with some correlation.
The ellipse oblateness and the rotation angle are calculated from the covariance matrix of the signal MC distribition after the event selection.
The center of the ellipse is the mean of the distribution.
The ellipse size is determined to maximize the figure-of-merit (FOM)~\cite{Punzi:2003bu},
\begin{equation}
FOM = \frac{\varepsilon}{\frac{\alpha}{2} + \sqrt{N_{\mathrm{BG}}}},
\label{eq:FigureOfMerit}
\end{equation}
where $\varepsilon$ is the signal efficiency inside the ellipse, $\alpha$ is the confidence coefficient ($\alpha =$ 1.64 at 90\% C.L.).

\begin{figure}[ht]
  \centering
  \begin{minipage}[b]{0.5\hsize}
    \centering
    \includegraphics[width=6.5cm,clip]{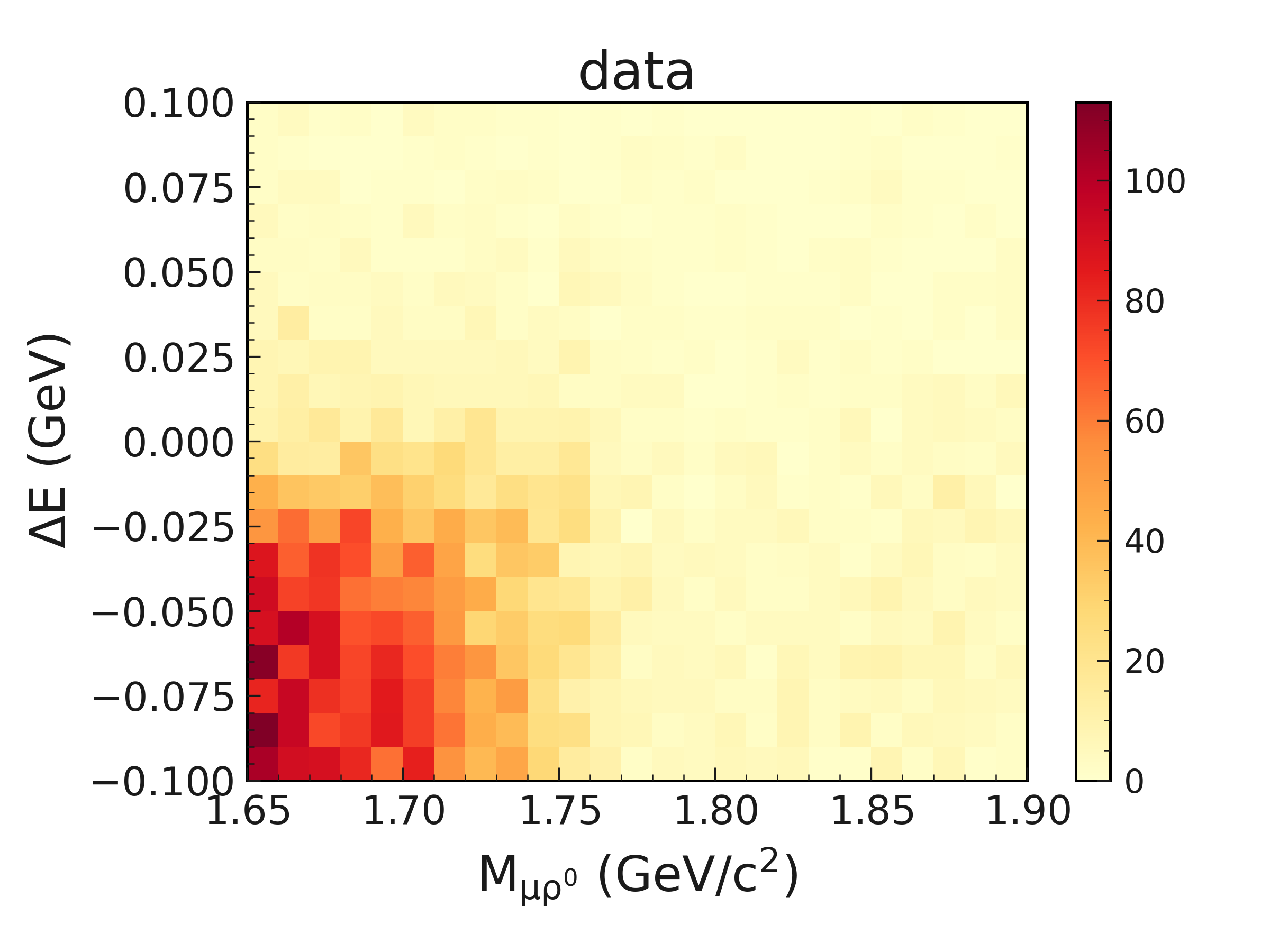}
  \end{minipage}\\
  \begin{minipage}[b]{0.45\hsize}
    \centering
    \includegraphics[width=6.5cm,clip]{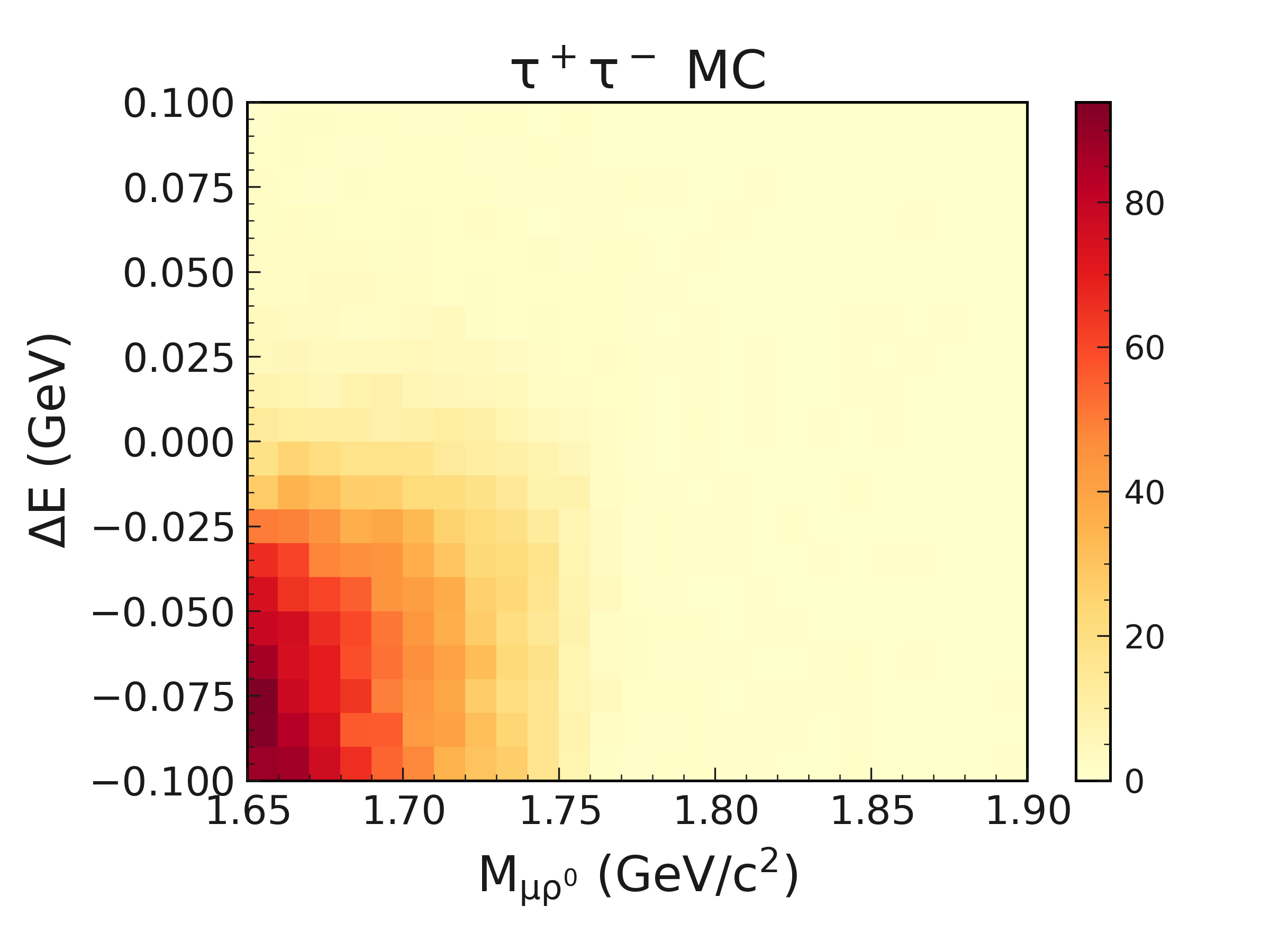}
  \end{minipage}
  \begin{minipage}[b]{0.45\hsize}
    \centering
    \includegraphics[width=6.5cm,clip]{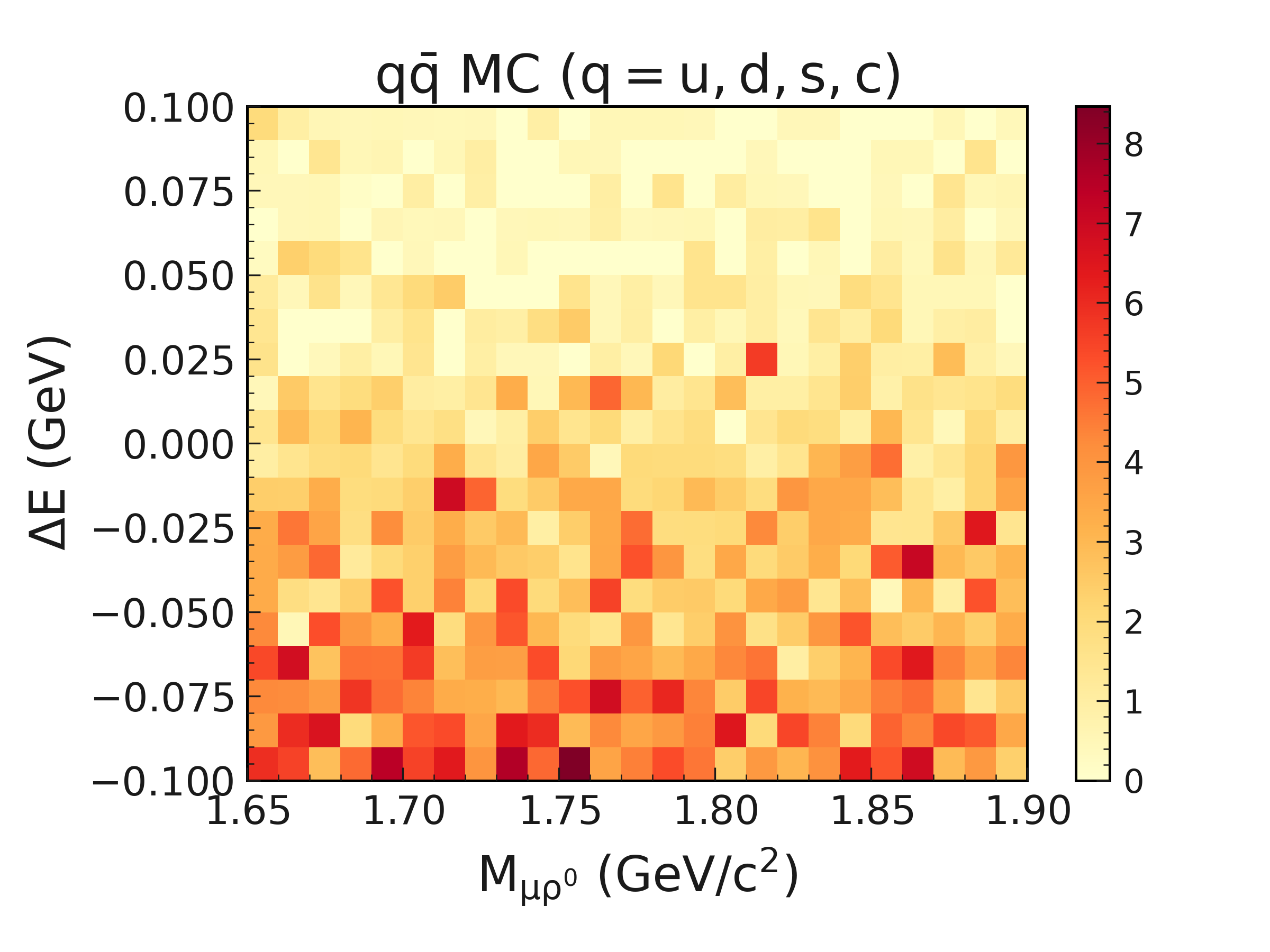}
  \end{minipage}
  \caption{The $M_{\ell V^0}$ vs. $\Delta E$ distribution of the $\tau \rightarrow \mu \rho^0$ hadron background enhanced samples: the data (upper side), the generic $\tau^+\tau^-$ MC (lower left) and the $q\bar{q}$ continuum MC (lower right, $q=u,d,s,c$). The range of the $\Delta E$ axis is limited to the fitting region. The MC sets are scaled to the data. The low-multiplicity background events are negligible for the hadron background enhanced samples and are not shown in this figure.}
  \label{fig:PIDsideband-mrho-data-mc}
\end{figure}

We estimate $N_{\mathrm{BG}}$ through interpolation from the sideband data.
Here the sideband data is a set of data passing the event selection and inside the sideband region: 1.65 GeV/$c^2$ $< M_{\ell V^0} <$ 1.9 GeV/$c^2$ and $|\Delta E| <$ 0.1 GeV outside of the blind region.
The interpolation is based on a function in the $M_{\ell V^0}$--$\Delta E$ plane.
This function is obtained by fitting the distribution of the hadron background enhanced data within $|\Delta E| <$ 0.1 GeV, and then it is scaled to the sideband data.
Figure \ref{fig:PIDsideband-mrho-data-mc} shows the distributions of the hadron background enhanced data and MC for the $\mu \rho^0$ mode.
The function is:
\begin{equation}
  F(M_{\ell V^0}, \Delta E) = f(M_{\ell V^0}) \times \frac{1}{1+\mathrm{exp}[a_y(\Delta E - y_0)]}+c_0^{\mathrm{flat}},
  \label{eq:2Dfit-function}
\end{equation}
\begin{equation}
  \begin{split}
    f(x) &= \left\{ \begin{array}{l}
      \displaystyle \int_{x-5\sigma}^{x+5\sigma} g(x^{\prime}) \times \frac{1}{\sqrt{2\pi}\sigma} \mathrm{exp}\Bigl[\frac{-(x-x^{\prime})^2}{2\sigma^2}\Bigl] \mathrm{d}x^{\prime} \;\;\; (V^0=\rho^0, \omega) \\
      c_1(x-x_0)^2 + c_0\;\;\; (V^0=K^{\ast0}, \overline{K}{}^{\ast0}) \\
      c_0 \;\;\; (V^0=\phi)
    \end{array} \right.\\
    &g(x) = \left\{ \begin{array}{l}
      \displaystyle c_1[(x-x_0)^2+k(x-x_0)] + c_0\;\;\; (x < x_0, V^0=\rho^0) \\
      c_1(x-x_0) + c_0\;\;\; (x < x_0, V^0=\omega) \\
      c_0 \;\;\; (x \geq x_0)
    \end{array} \right.
  \end{split}
  \label{eq:M_tau-fit-function}
\end{equation}
where $f(x)$ represents the background distribution as a function of $M_{\ell V^0}$; $c_1$, $c_0$, $x_0$, and $k$ are parameters that define the shape of the function; $a_y$ represents sharpness of the sigmoid function along the $\Delta E$ axis; $y_0$ is the center of the sigmoid function; and $c_0^{\mathrm{flat}}$ is a term of flat background events in the $M_{\ell V^0}$--$\Delta E$ plane.
We define $f(x)$ for each $V^0$ in eq.~\eqref{eq:M_tau-fit-function} and the functions for the $\ell \rho^0$ ($\ell \omega$) modes are smeared by a Gaussian with standard deviation ($\sigma$) of 6.6 (9.6) MeV/c$^2$.
This $\sigma$ corresponds to the mass resolution that affects the edge of the $M_{\ell V^0}$ distribution close to the $\tau$ mass for the $\tau^+\tau^-$ background.
The edge is broad for the other modes owing to wrong mass assignment of fake kaons.
The $\tau^+ \tau^-$ background events for the $\ell \phi$ modes are included in $c_0$ because they are flat along the $M_{\ell V^0}$ axis in 1.65 GeV/$c^2$ $< M_{\ell V^0} <$ 1.9 GeV/$c^2$.

We obtain the optimal fit parameters by a likelihood fit using MINUIT~\cite{James:1975dr}.
The following region is excluded from the fitting to avoid $D^{+} \rightarrow K^-\pi^+\pi^+$ and $D^{+} \rightarrow \pi^+\phi$ background events, which cluster around the $D$ meson mass: 1.83(1.82) GeV/$c^2$ $\leq M_{\ell V^0} <$ 1.89 GeV/$c^2$ and $\Delta E <$ 0.04 GeV for the $\mu K^{\ast0}$ ($e K^{\ast0}$) and $\mu \phi$ ($e \phi$) modes.

The parameters of $a_y$, $y_0$, $k$, and $x_0$ are fixed at the fit results of the hadron background enhanced data within $|\Delta E| <$ 0.1 GeV.
The fit uncertainties of these fixed parameters are included in the systematic uncertainty of $N_{\mathrm{BG}}$.
The other fit parameters correspond to the scale factors of each background component: generic $\tau^+\tau^-$ ($c_1$), and continuum and low-multiplicity background events ($c_0$ and $c_0^{\mathrm{flat}}$).
We fit the function floating these scale factors ($c_1$, $c_0$, and $c_0^{\mathrm{flat}}$) to the sideband data.
The same region around the $D$ meson mass as for the fit to the hadron background enhanced data is excluded from the fitting for the $\ell \phi$ and $\ell K^{\ast0}$ modes.
The functions are integrated in the elliptical signal regions to deduce $N_{\mathrm{BG}}$, which are in the range of 0.25--0.95.

Other systematic uncertainties on $N_{\mathrm{BG}}$ come from the difference of the $M_{\ell V^0}$--$\Delta E$ distributions between the sideband data and the hadron background enhanced data within $|\Delta E| < $ 0.1 GeV.
The difference originates from the BDT selection and the lepton identification.
We evaluate the amounts of changes of $N_{\mathrm{BG}}$ when the parameters---$a_y$, $y_0$, $k$, and $x_0$---are redetermined with another $M_{\ell V^0}$--$\Delta E$ distribution of the hadron background enhanced data changed by the BDT selection or weighted for the lepton identification as described below.
The amounts of changes of $N_{\mathrm{BG}}$ are taken as the systematic uncertainties of $N_{\mathrm{BG}}$.

Because the 80\% of the hadron background enhanced data are used for the BDT training, we apply the BDT selection for the rest of data (the validation samples) and redetermine the parameters with those data.
We estimate $N_{\mathrm{BG}}$ in the same way as previously described, and the amount of change from the nominal $N_{\mathrm{BG}}$ is obtained.

Each event of the sideband data has a pion misidentified as a lepton, which tends to have a lower momentum than the pions in the hadron background enhanced data.
That is the reason of the difference in the $M_{\ell V^0}$--$\Delta E$ distributions because the fake rate, $R_{e(\mu)}^{\mathrm{fake}} (P, \theta)$, depends on the momentum $P$ and $\theta$ of the track.
We generate weighted hadron background enhanced data, 
where each event is weighted by the ratio of $R_{e(\mu)}^{\mathrm{fake}} (P, \theta)$ to $1 - R_{e}^{\mathrm{fake}} (P, \theta) - R_{\mu}^{\mathrm{fake}} (P, \theta)$ for the track, in order to conform the $M_{\ell V^0}$--$\Delta E$ distribution to the one of the sideband data.
We redetermine the parameters with this weighted data and evaluate the change of $N_{\mathrm{BG}}$ again.

The statistical uncertainty of $N_{\mathrm{BG}}$ is calculated as follows:
We generate 100 sets of pseudo-data for each mode in the $M_{\ell V^0}$--$\Delta E$ histogram.
The content of each bin in the histogram is set randomly following a Poisson distribution, with the mean taken from the function fitted to the sideband data.
We fit the function to each set of the pseudo-data to deduce $N_{\mathrm{BG}}$, and the standard deviation of these $N_{\mathrm{BG}}$ is taken as the statistical uncertainty.

The major contribution to $N_{\mathrm{BG}}$ comes from the $M_{\ell V^0}$ flat term in eq.~\eqref{eq:2Dfit-function} ($c_0$ and $c_0^{\mathrm{flat}}$), which corresponds to the continuum or low-multiplicity background events.
The contribution of the generic $\tau^+\tau^-$ background events, which depends on $M_{\ell V^0}$, is about one-third as large as the other background contributions.
We cannot distinguish the background components of the $\ell \phi$ modes through the fit to the data, because the generic $\tau^+\tau^-$ background events are distributed evenly along the $M_{\ell V^0}$ axis.

The systematic uncertainties of the expected number of signal events are listed in Table~\ref{tbl:Uncertainty-Summary}.
The dominant uncertainties are from the particle identification.

The track and photon energy resolutions in the MC are corrected such that the mass resolution of the $D^{(\ast)+}$ meson matches between the data and MC, where $D^{(\ast)+} \rightarrow K^-\pi^+\pi^+(\pi^0)$ is reconstructed with similar event selection criteria to the signal ones (e.g. $|\Delta E| <$ 0.5 GeV).
The uncertainty of the data mass resolution propagates to the uncertainties of the corrected energy resolutions.
We generate two additional signal MC sets in which the track (photon) energy resolution is different by plus and minus its uncertainty, and take the half of the difference in the expected number of the signal events as the systematic uncertainty.

All the uncertainties in Table~\ref{tbl:Uncertainty-Summary} are summed in quadrature to yield the total systematic uncertainties shown in Table~\ref{tbl:SummaryTable}.

\begin{table*}[htb]
  \centering
  \caption{List of the systematic uncertainties of the expected number of signal events. The average number of tracks (particles) in the reconstructed $\tau^+\tau^-$ events for each signal mode is represented as $N_{\mathrm{track (particle)}}$. When the uncertainty is different mode by mode, we show the range of the uncertainty.}
  \label{tbl:Uncertainty-Summary}
  \begin{tabular}{|c|c|}
    \hline
    Source  & $\sigma_{\mathrm{syst}}$ (\%)\\
    \hline
    Integrated luminosity & 1.4 \\
    $ee\rightarrow\tau\tau(\gamma)$ cross section~\cite{Banerjee:2007is} & 0.3 \\
    $\mathcal{B}$($\phi \rightarrow K^+K^-$) and $\mathcal{B}$($\omega \rightarrow \pi^+ \pi^- \pi^0$) & 1.2 and 0.7 \\
    Trigger efficiency & 0.2--0.9 \\
    Tracking efficiency & $0.35 \times N_{\mathrm{track}}$ \\
    Electron identification efficiency & $1.7 \times N_{\mathrm{electron}}$ \\
    Muon identification efficiency & $1.8 \times N_{\mathrm{muon}}$ \\
    $K^{\pm}$ and $\pi^{\pm}$ identification efficiency & 1.6 ($\rho^0$), 1.8 ($\phi$) and 1.1 ($K^{\ast0}$ and $\overline{K}{}^{\ast0}$) \\
    $\pi^0$ efficiency & $2.2 \times N_{\pi^0}$ \\
    Electron veto for hadrons & 0.4--1.2 \\
    MC statistics & 0.3--0.5 \\
    Track energy resolution & 0.3--1.3 \\
    Photon energy resolution & 0.0--0.4 \\
    \hline
  \end{tabular}
\end{table*}

\section{Results}
\label{sec:limits}

\begin{figure}[ht]
  \centering
  \begin{minipage}[b]{0.49\hsize}
    \centering
    \includegraphics[width=7.5cm,clip]{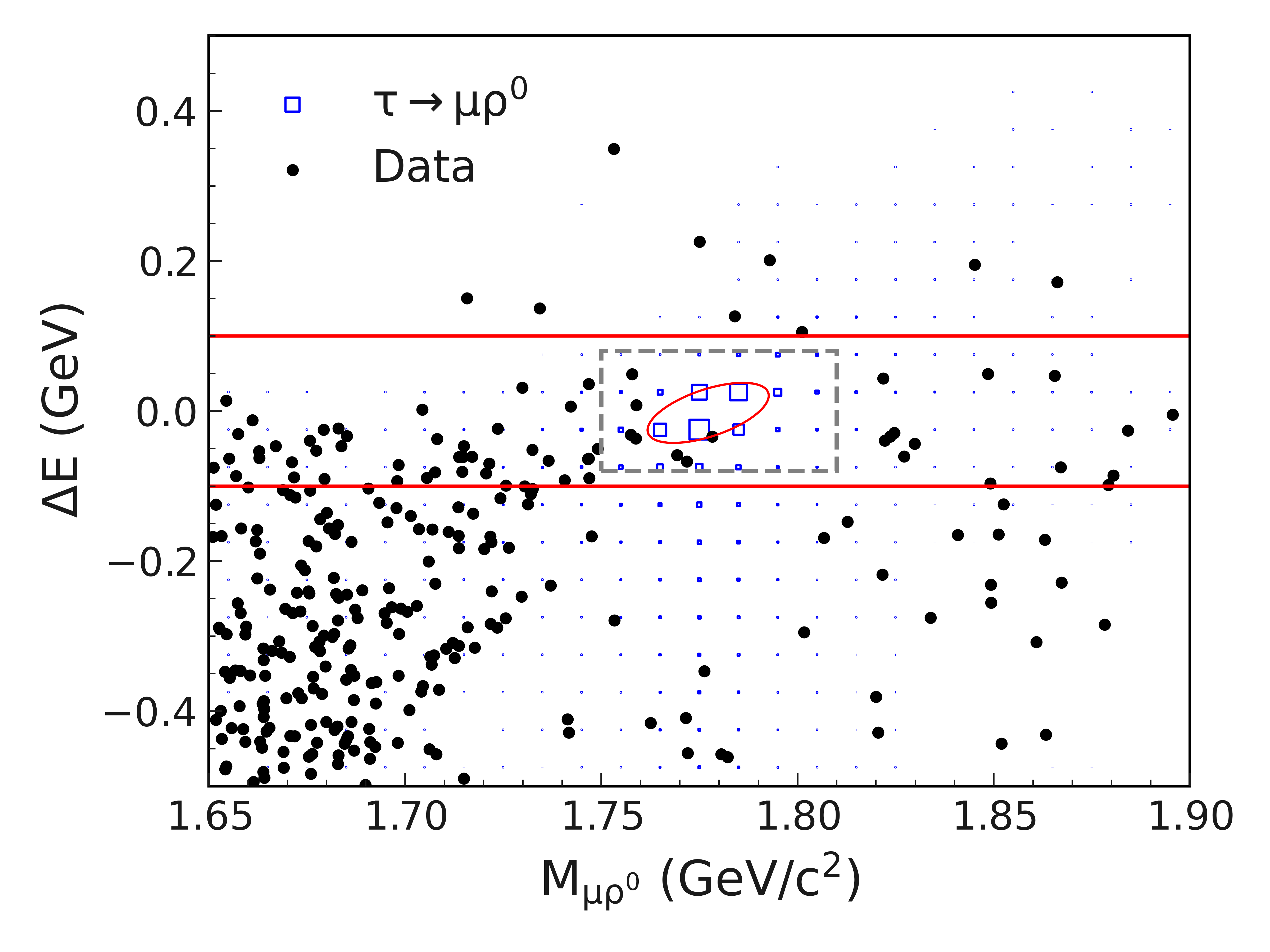}
    \subcaption{$\tau \rightarrow \mu \rho^0$}
  \end{minipage}
  \begin{minipage}[b]{0.49\hsize}
    \centering
    \includegraphics[width=7.5cm,clip]{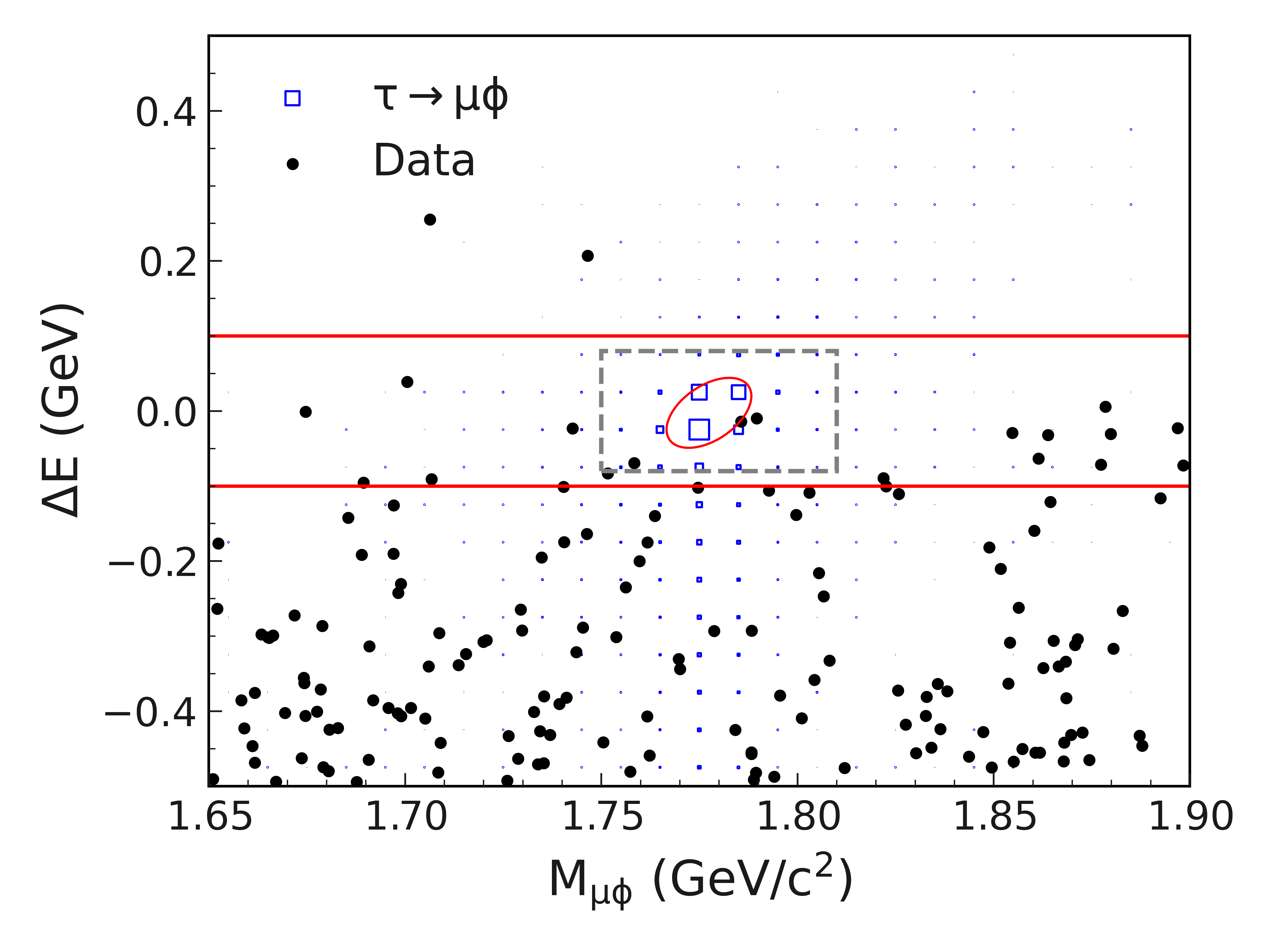}
    \subcaption{$\tau \rightarrow \mu \phi$}
  \end{minipage}\\
  \begin{minipage}[b]{0.49\hsize}
    \centering
    \includegraphics[width=7.5cm,clip]{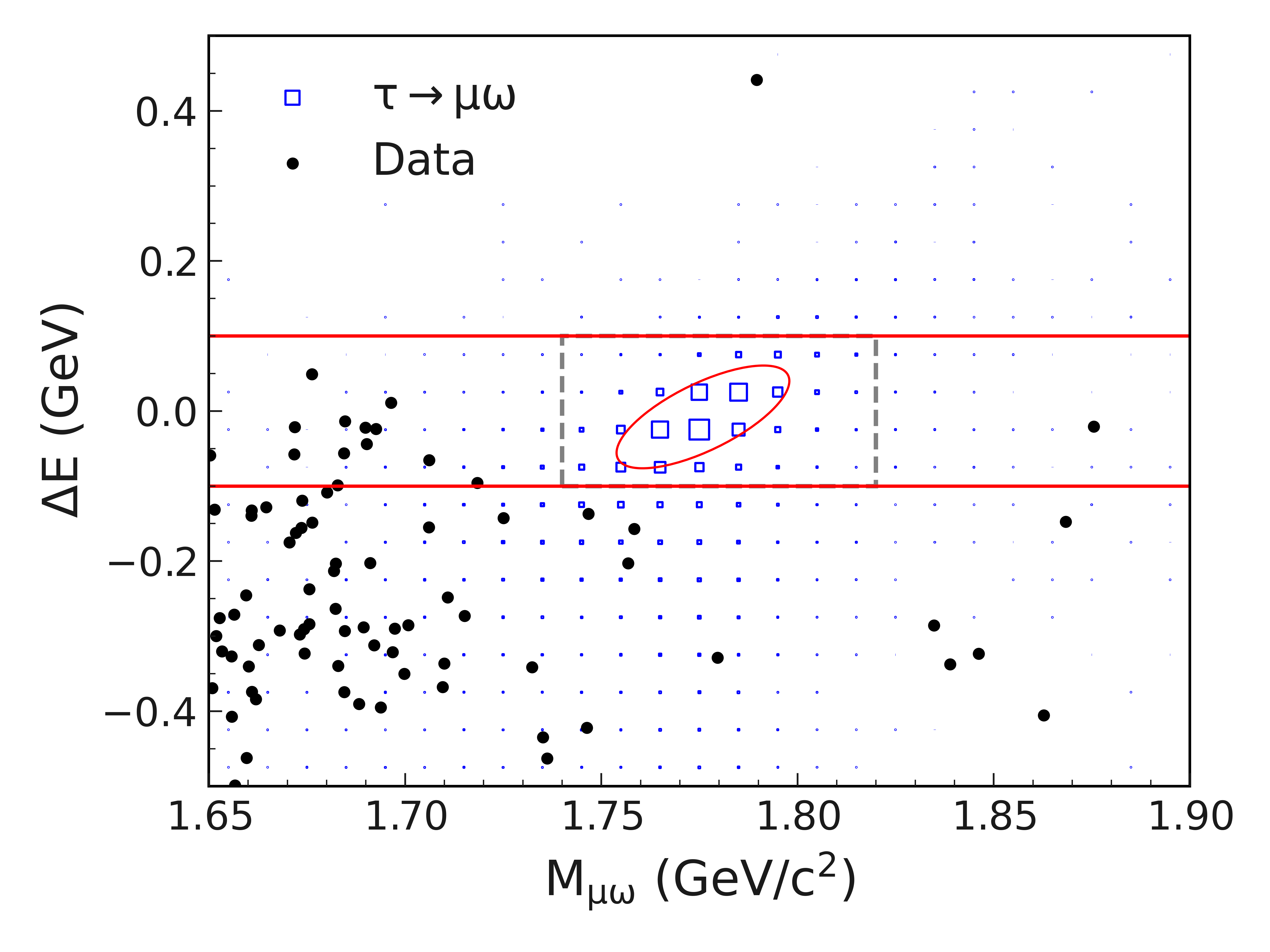}
    \subcaption{$\tau \rightarrow \mu \omega$}
  \end{minipage}
  \begin{minipage}[b]{0.49\hsize}
    \centering
    \includegraphics[width=7.5cm,clip]{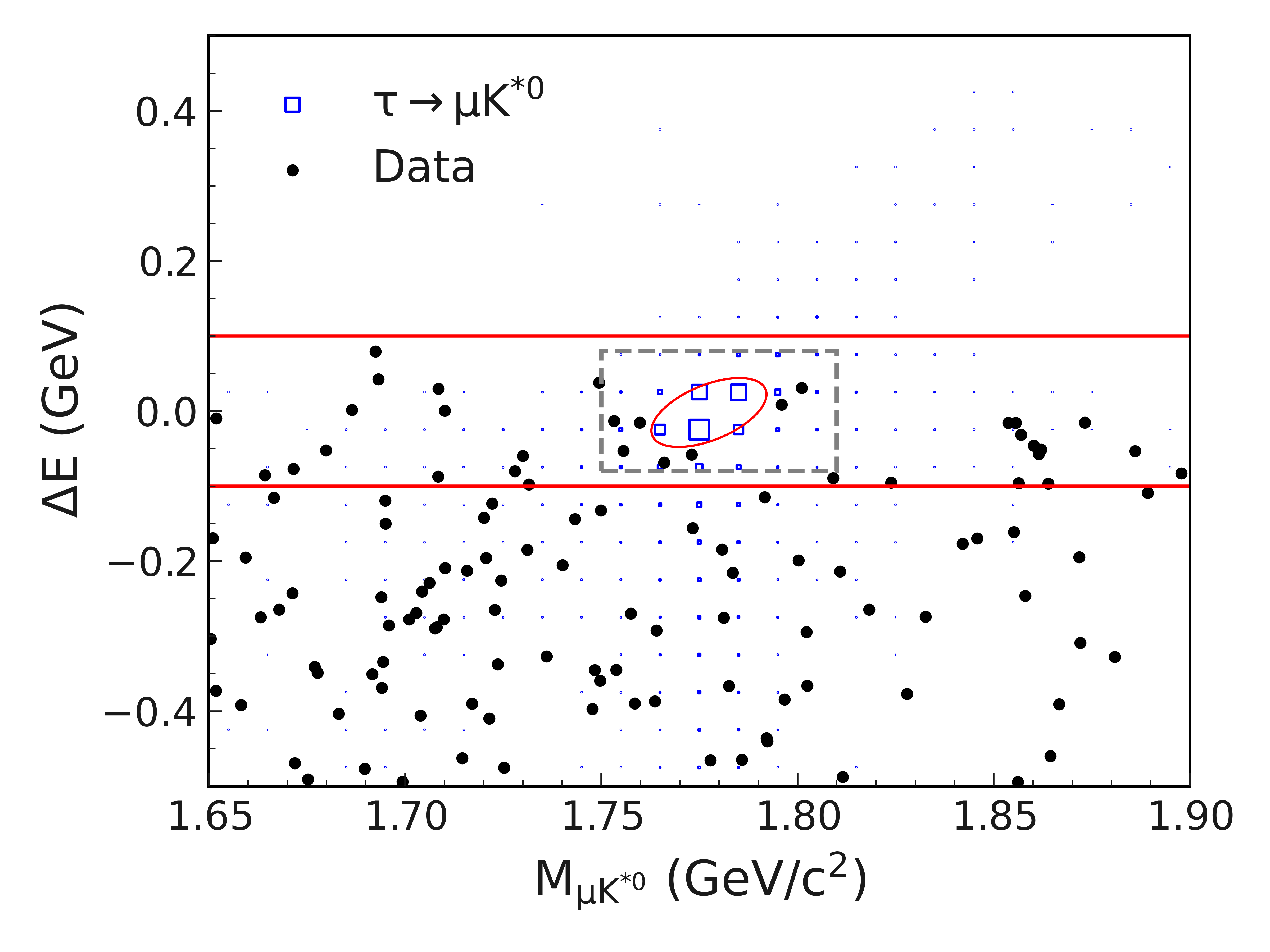}
    \subcaption{$\tau \rightarrow \mu K^{\ast0}$}
  \end{minipage}\\
  \begin{minipage}[b]{0.49\hsize}
    \centering
    \includegraphics[width=7.5cm,clip]{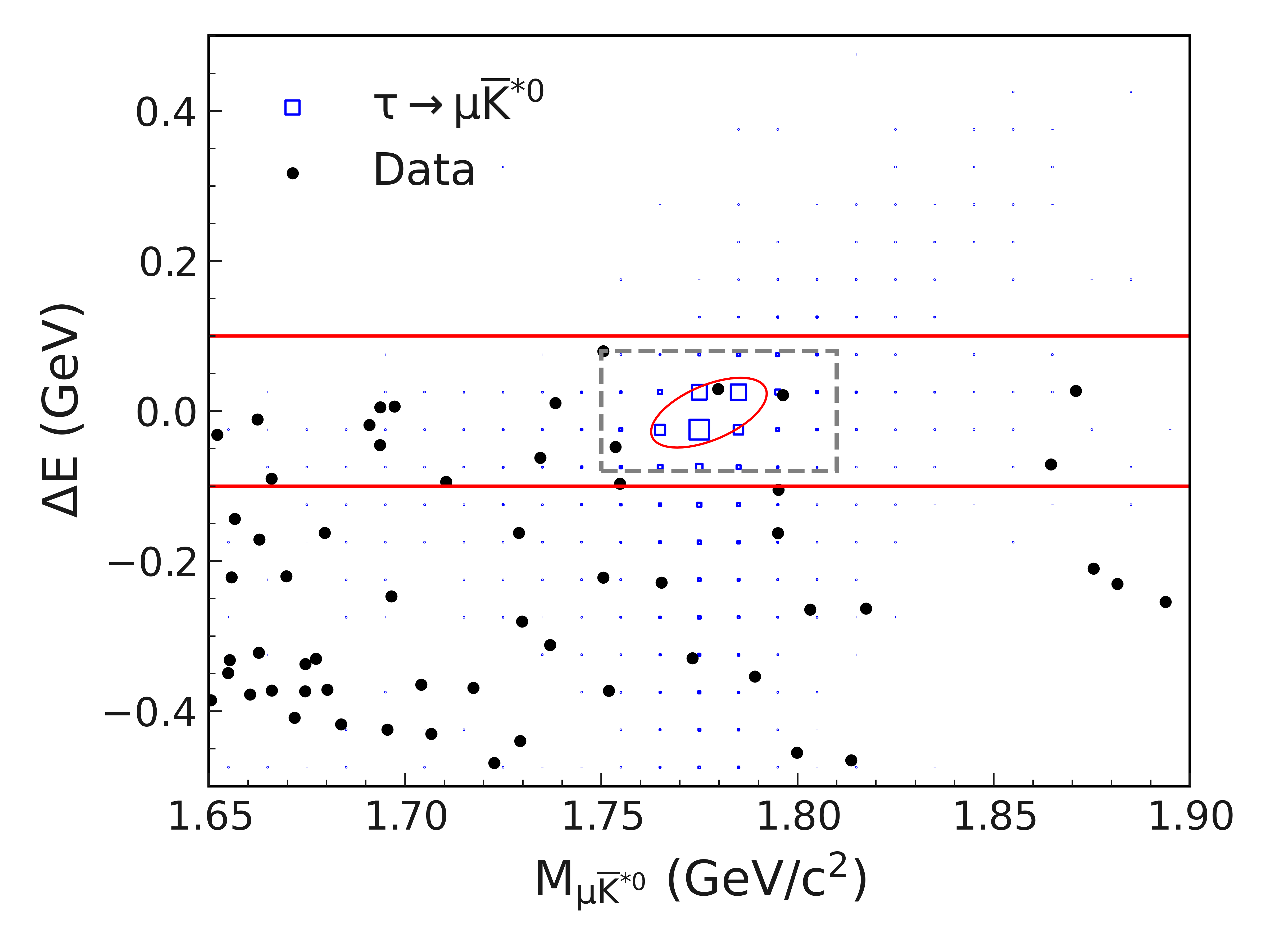}
    \subcaption{$\tau \rightarrow \mu \overline{K}{}^{\ast0}$}
  \end{minipage}
  \caption{Observed event distributions of $M_{\ell V^0}$ vs. $\Delta E$ after the $\tau \rightarrow \mu V^0$ event selection. Black points are the data, blue squares show the signal MC distribution with an arbitrary normalization. The red elliptical lines are the signal regions. The estimations of the number of background events are done using the data between the red horizontal lines outside the blind regions (the gray dashed rectangles).}
  \label{fig:unblind-BDT-mV0}
\end{figure}

\begin{figure}[ht]
  \centering
  \begin{minipage}[b]{0.49\hsize}
    \centering
    \includegraphics[width=7.5cm,clip]{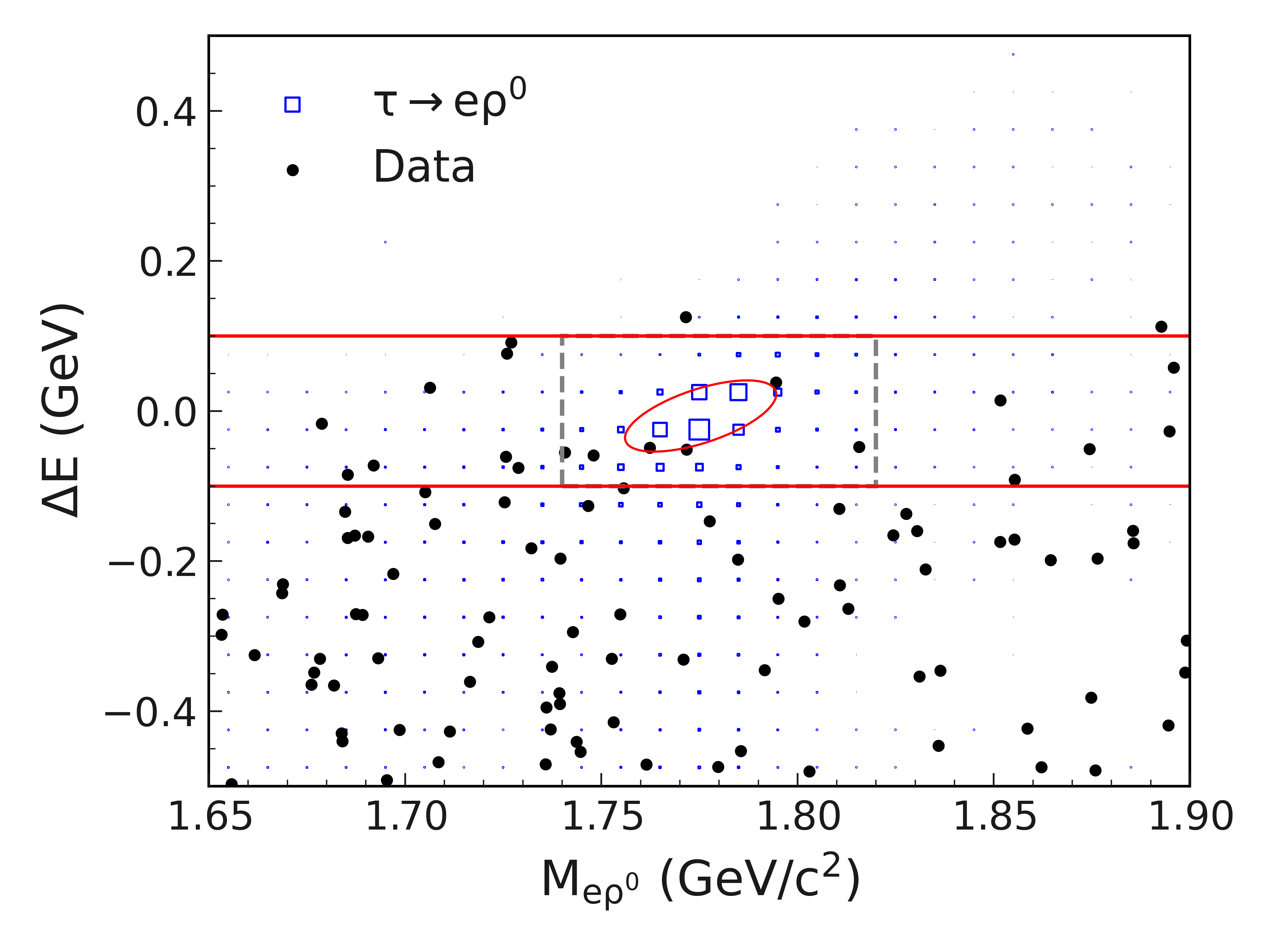}
    \subcaption{$\tau \rightarrow e \rho^0$}
  \end{minipage}
  \begin{minipage}[b]{0.49\hsize}
    \centering
    \includegraphics[width=7.5cm,clip]{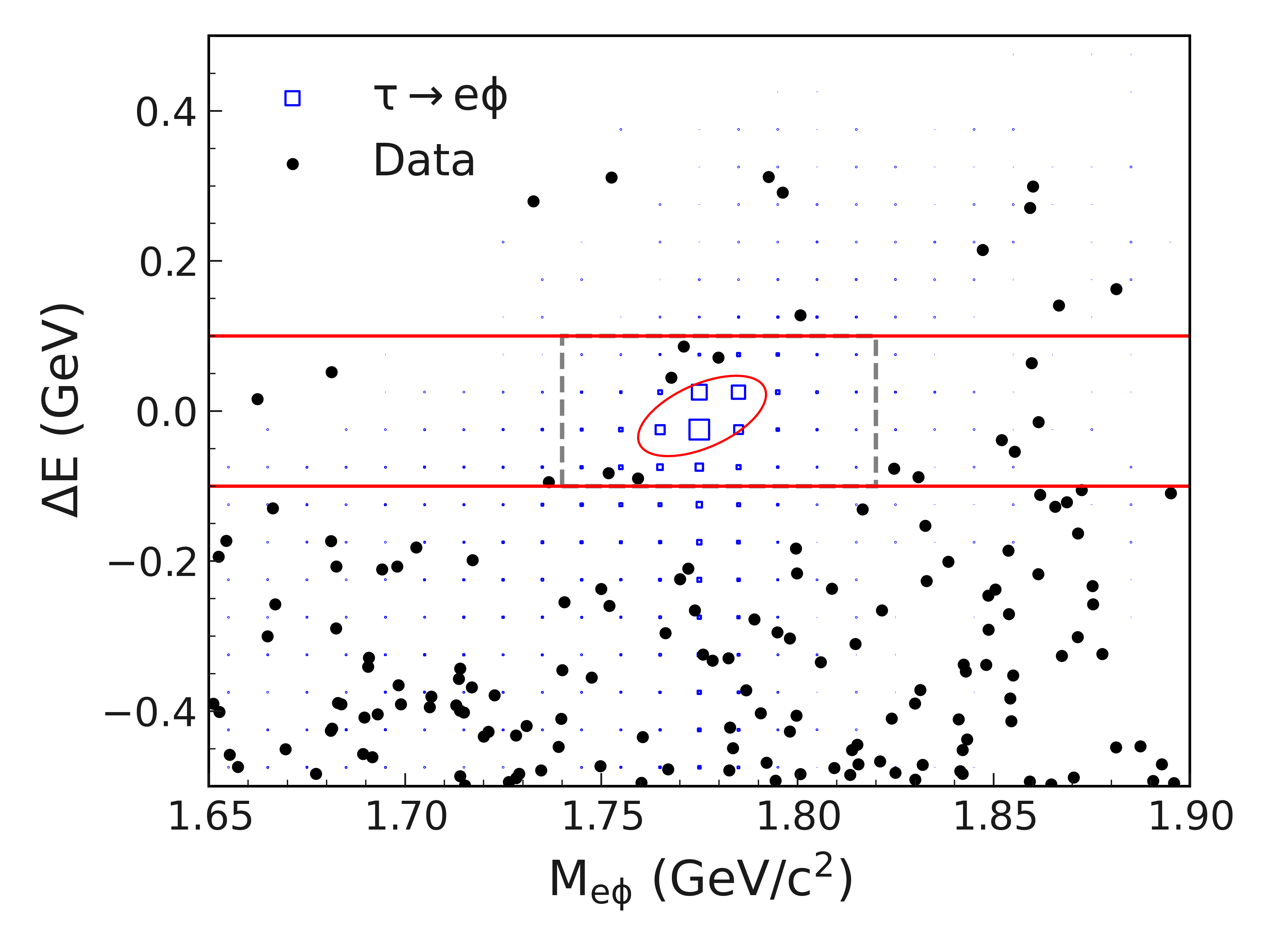}
    \subcaption{$\tau \rightarrow e \phi$}
  \end{minipage}\\
  \begin{minipage}[b]{0.49\hsize}
    \centering
    \includegraphics[width=7.5cm,clip]{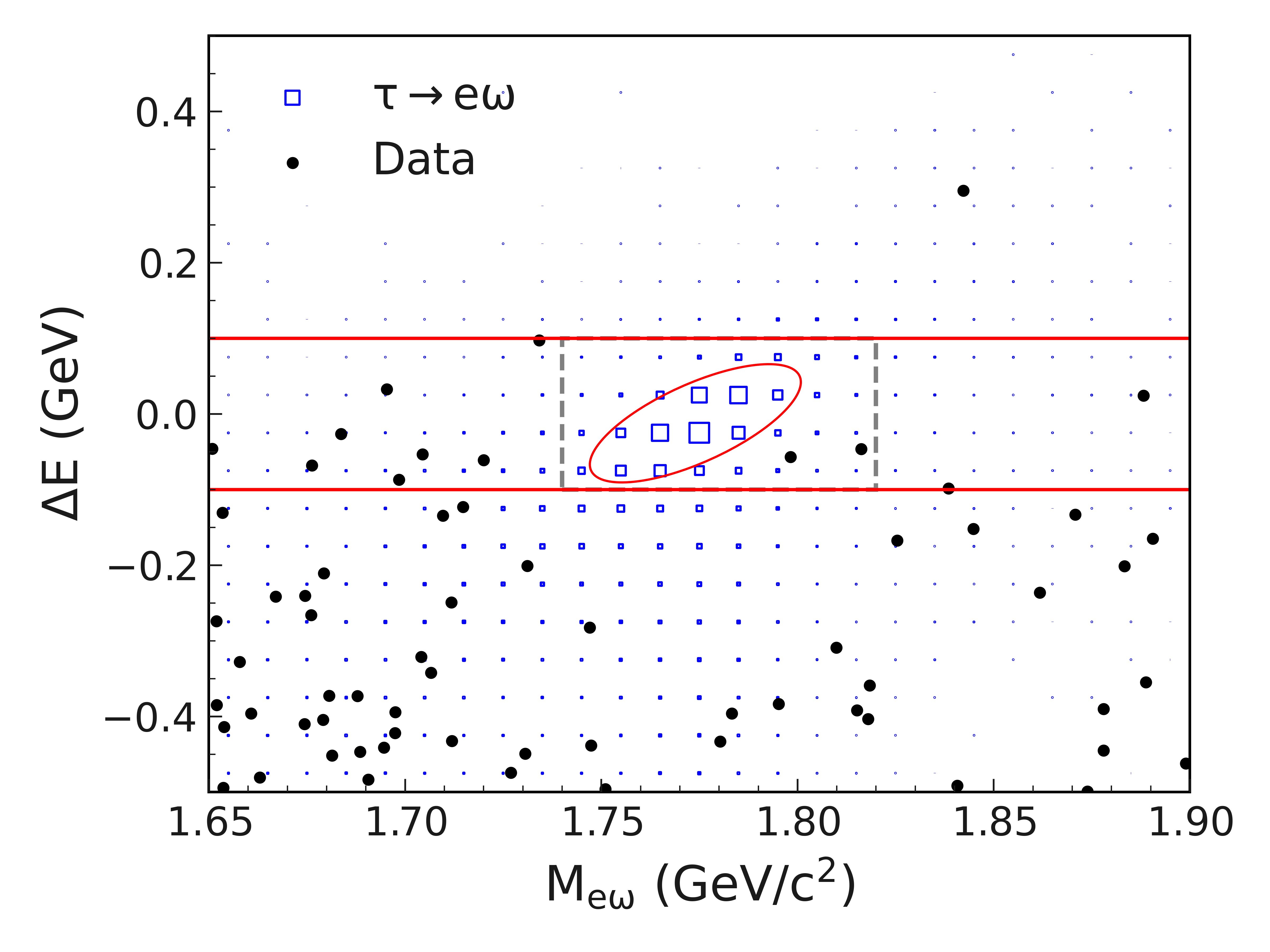}
    \subcaption{$\tau \rightarrow e \omega$}
  \end{minipage}
  \begin{minipage}[b]{0.49\hsize}
    \centering
    \includegraphics[width=7.5cm,clip]{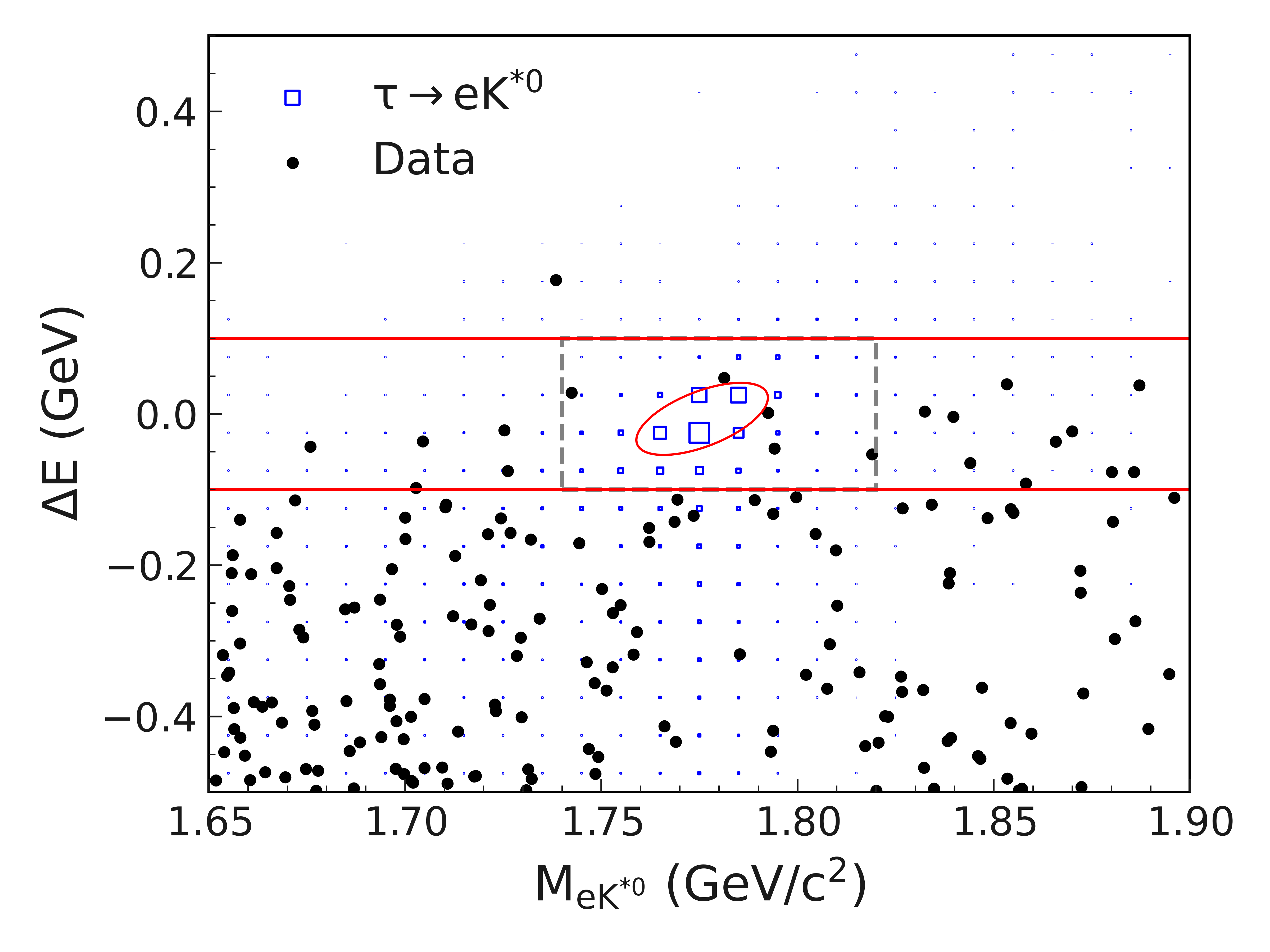}
    \subcaption{$\tau \rightarrow e K^{\ast0}$}
  \end{minipage}\\
  \begin{minipage}[b]{0.49\hsize}
    \centering
    \includegraphics[width=7.5cm,clip]{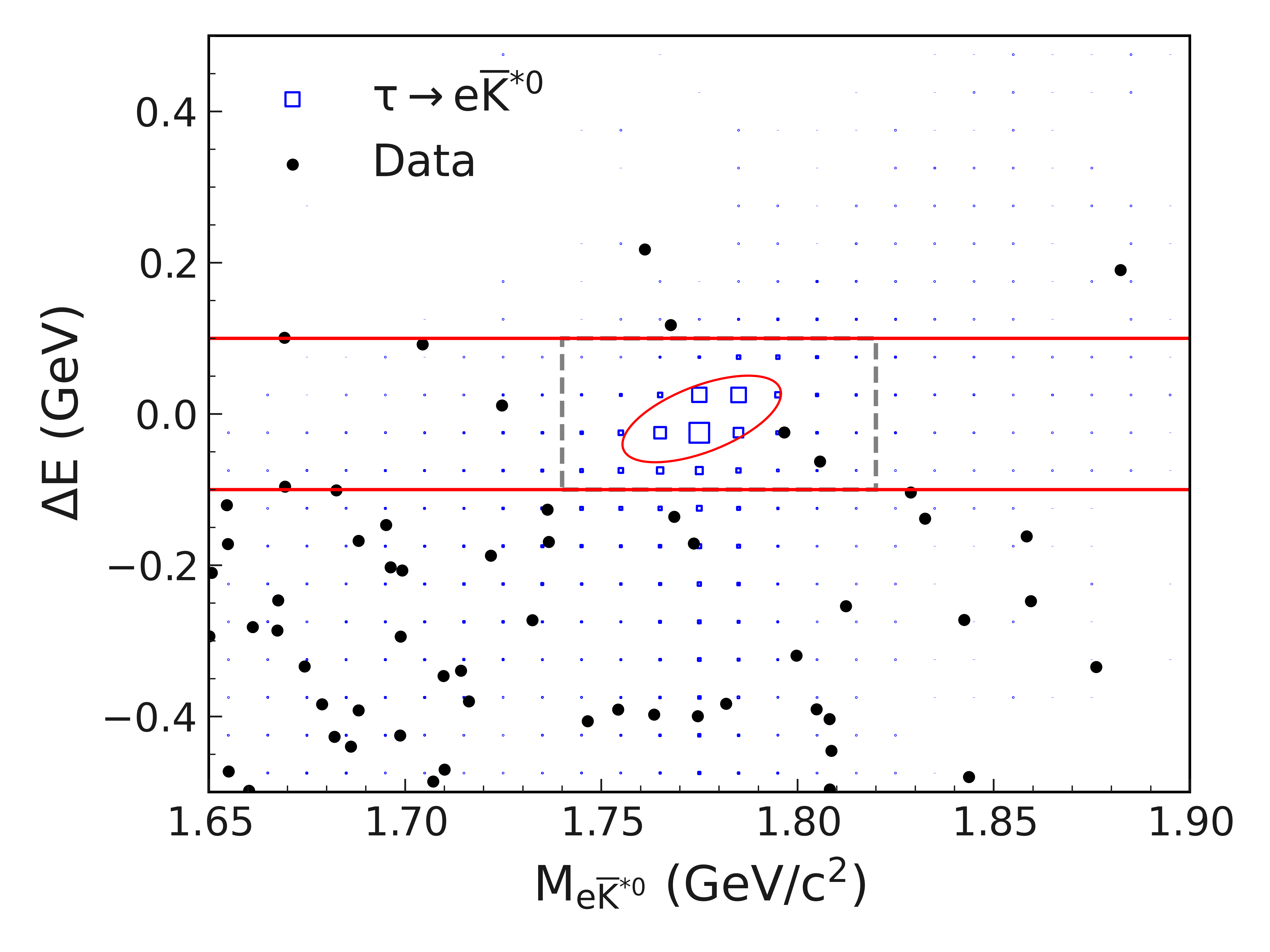}
    \subcaption{$\tau \rightarrow e \overline{K}{}^{\ast0}$}
  \end{minipage}
  \caption{Observed event distributions of $M_{\ell V^0}$ vs. $\Delta E$ after the $\tau \rightarrow e V^0$ event selection. Black points are the data, blue squares show the signal MC distribution with an arbitrary normalization. The red elliptical lines are the signal regions. The estimations of the number of background events are done using the data between the red horizontal lines outside the blind regions (the gray dashed rectangles).}
  \label{fig:unblind-BDT-eV0}
\end{figure}

Figures \ref{fig:unblind-BDT-mV0} and \ref{fig:unblind-BDT-eV0} show the observed event distributions in the $M_{\ell V^0}$--$\Delta E$ plane.
The observed number of events in the signal region ($N_{\mathrm{obs}}$) has no excess over $N_{\mathrm{BG}}$.

We set 90\% C.L. upper limits on the branching fractions based on a Bayesian method with the use of Markov Chain Monte Carlo~\cite{Caldwell:2008fw}.
The probability density function of the branching fraction ($\mathcal{B}$($\tau \rightarrow \ell V^0$)) is calculated assuming that $N_{\mathrm{obs}}$ follows a Poisson distribution function whose mean value is the expected number of events ($N_{\mathrm{exp}}$),
\begin{equation}
N_{\mathrm{exp}} = L\times 2\sigma_{\tau\tau}\mathcal{B}(\tau \rightarrow \ell V^0) \times \varepsilon + N_{\mathrm{BG}},
\label{eq:expectedNumberOfEvents}
\end{equation}
where $L$ is the integrated luminosity ($980.4 \pm 13.7\ \mathrm{fb}^{-1}$), $\sigma_{\tau\tau}$ is the cross section of $\tau$-pair production that is calculated with KKMC~\cite{Banerjee:2007is} (the weighted average of $\sigma_{\tau\tau}$ at all the beam energies is $0.916\pm 0.003$ nb), and $\varepsilon$ is the signal efficiency including the branching fraction of the $V^0$.
We assume that these values ($L, \sigma_{\tau\tau}, \varepsilon$, and $N_{\mathrm{BG}}$) follow a Gaussian distribution with the width equal to the uncertainty of each value.

The upper limits on $\mathcal{B}$($\tau \rightarrow \ell V^0$) are listed in Table~\ref{tbl:SummaryTable}.
The average of the limits is better than that of the previous results using 854 fb$^{-1}$~\cite{Belle:2011ogy} by 30\%.
This is due to the additional 15\% of integrated luminosity; the addition of $\pi^{\pm}\pi^{\mp}\pi^{\pm}\nu$ and $\pi^{\pm}\pi^0\pi^0\nu$ modes in $\tau_{\mathrm{tag}}$ reconstruction, which increases the signal efficiency by 9.6\%; and the event selection by multivariate analysis (BDT).
The upper limit on $\mathcal{B}$($\tau \rightarrow \mu \rho^0$) is worse than that of the previous result, though the expected upper limit before unblinding is better.
This is because we use the Bayesian limits instead of the Frequentist limits, which are negatively proportional to $N_{\mathrm{BG}}$ when $N_{\mathrm{obs}}$ is fixed.

\begin{table*}[h]
  \centering
  \caption{The signal efficiency ($\varepsilon$), the expected number of background events ($N_{\mathrm{BG}}$), total systematic uncertainty of the expected number of signal events ($\sigma_{\mathrm{syst}}$), the number of observed events in the signal region ($N_{\mathrm{obs}}$), and the observed 90\% C.L. upper limits on the branching fraction ($\mathcal{B}_{\mathrm{obs}}$ (10$^{-8}$)).}
  \label{tbl:SummaryTable}
  \begin{tabular}{cccccc}
    \hline
    \hline
    Mode  & $\varepsilon$ (\%) & $N_{\mathrm{BG}}$ & $\sigma_{\mathrm{syst}}$ (\%) & $N_{\mathrm{obs}}$ & $\mathcal{B}_{\mathrm{obs}}$ ($\times 10^{-8}$) \\
    \hline
    $\tau^\pm \rightarrow \mu^\pm \rho^0     $ & 7.78 & 0.95$\pm$0.20(stat.) $\pm$0.15(syst.) & 4.6 & 0 & $<$ 1.7 \\
    $\tau^\pm \rightarrow e^\pm   \rho^0     $ & 8.49 & 0.80$\pm$0.27(stat.) $\pm$0.04(syst.) & 4.4 & 1 & $<$ 2.2 \\
    $\tau^\pm \rightarrow \mu^\pm \phi       $ & 5.59 & 0.47$\pm$0.15(stat.) $\pm$0.05(syst.) & 4.8 & 0 & $<$ 2.3 \\
    $\tau^\pm \rightarrow e^\pm   \phi       $ & 6.45 & 0.38$\pm$0.21(stat.) $\pm$0.00(syst.) & 4.5 & 0 & $<$ 2.0 \\
    $\tau^\pm \rightarrow \mu^\pm \omega     $ & 3.27 & 0.32$\pm$0.23(stat.) $\pm$0.19(syst.) & 4.8 & 0 & $<$ 3.9 \\
    $\tau^\pm \rightarrow e^\pm   \omega     $ & 5.41 & 0.74$\pm$0.43(stat.) $\pm$0.06(syst.) & 4.5 & 0 & $<$ 2.4 \\
    $\tau^\pm \rightarrow \mu^\pm K^{\ast0}      $ & 4.52 & 0.84$\pm$0.25(stat.) $\pm$0.31(syst.) & 4.3 & 0 & $<$ 2.9 \\
    $\tau^\pm \rightarrow e^\pm   K^{\ast0}      $ & 6.94 & 0.54$\pm$0.21(stat.) $\pm$0.16(syst.) & 4.1 & 0 & $<$ 1.9 \\
    $\tau^\pm \rightarrow \mu^\pm \overline{K}{}^{\ast0}$ & 4.58 & 0.58$\pm$0.17(stat.) $\pm$0.12(syst.) & 4.3 & 1 & $<$ 4.3 \\
    $\tau^\pm \rightarrow e^\pm   \overline{K}{}^{\ast0}$ & 7.45 & 0.25$\pm$0.11(stat.) $\pm$0.02(syst.) & 4.1 & 0 & $<$ 1.7 \\
    \hline
    \hline
  \end{tabular}
\end{table*}

\section{Conclusion}
\label{sec:conclusion}
To conclude, we searched for lepton-flavor-violating $\tau$ decays into one lepton and one vector meson using the full 980 fb$^{-1}$ of Belle data.
No statistically significant signal candidates are observed, and the 90\% C.L. upper limits on the branching fraction are in the range of (1.7--$4.3) \times 10^{-8}$ for $\tau \rightarrow \mu V^0$ and (1.7--$2.4) \times 10^{-8}$ for $\tau \rightarrow e V^0$.
The upper limits are improved by 30\% on average from the previous results.
We achieve these improvements both with the reconsideration of the event selection criteria and with the 126 fb$^{-1}$ of additional data set.

\acknowledgments

This work, based on data collected using the Belle detector, which was
operated until June 2010, was supported by 
the Ministry of Education, Culture, Sports, Science, and
Technology (MEXT) of Japan, the Japan Society for the 
Promotion of Science (JSPS), in particular the Grant-in-Aid for Scientific Research (S) 18H05226 and (A) 19H00682, and the Tau-Lepton Physics 
Research Center of Nagoya University; 
the Australian Research Council including grants
DP180102629, 
DP170102389, 
DP170102204, 
DE220100462, 
DP150103061, 
FT130100303; 
Austrian Federal Ministry of Education, Science and Research (FWF) and
FWF Austrian Science Fund No.~P~31361-N36;
the National Natural Science Foundation of China under Contracts
No.~11675166,  
No.~11705209;  
No.~11975076;  
No.~12135005;  
No.~12175041;  
No.~12161141008; 
Key Research Program of Frontier Sciences, Chinese Academy of Sciences (CAS), Grant No.~QYZDJ-SSW-SLH011; 
Project ZR2022JQ02 supported by Shandong Provincial Natural Science Foundation;
the Ministry of Education, Youth and Sports of the Czech
Republic under Contract No.~LTT17020;
the Czech Science Foundation Grant No. 22-18469S;
Horizon 2020 ERC Advanced Grant No.~884719 and ERC Starting Grant No.~947006 ``InterLeptons'' (European Union);
the Carl Zeiss Foundation, the Deutsche Forschungsgemeinschaft, the
Excellence Cluster Universe, and the VolkswagenStiftung;
the Department of Atomic Energy (Project Identification No. RTI 4002) and the Department of Science and Technology of India; 
the Istituto Nazionale di Fisica Nucleare of Italy; 
National Research Foundation (NRF) of Korea Grant
Nos.~2016R1\-D1A1B\-02012900, 2018R1\-A2B\-3003643,
2018R1\-A6A1A\-06024970, RS\-2022\-00197659,
2019R1\-I1A3A\-01058933, 2021R1\-A6A1A\-03043957,
2021R1\-F1A\-1060423, 2021R1\-F1A\-1064008, 2021R1\-A4A\-2001897, 2022R1\-A2C\-1003993;
Radiation Science Research Institute, Foreign Large-size Research Facility Application Supporting project, the Global Science Experimental Data Hub Center of the Korea Institute of Science and Technology Information and KREONET/GLORIAD;
the Polish Ministry of Science and Higher Education and 
the National Science Center;
the Ministry of Science and Higher Education of the Russian Federation, Agreement 14.W03.31.0026, 
and the HSE University Basic Research Program, Moscow; 
University of Tabuk research grants
S-1440-0321, S-0256-1438, and S-0280-1439 (Saudi Arabia);
the Slovenian Research Agency Grant Nos. J1-9124 and P1-0135;
Ikerbasque, Basque Foundation for Science, Spain;
the Swiss National Science Foundation; 
the Ministry of Education and the Ministry of Science and Technology of Taiwan;
and the United States Department of Energy and the National Science Foundation.
These acknowledgements are not to be interpreted as an endorsement of any
statement made by any of our institutes, funding agencies, governments, or
their representatives.
We thank the KEKB group for the excellent operation of the
accelerator; the KEK cryogenics group for the efficient
operation of the solenoid; and the KEK computer group and the Pacific Northwest National
Laboratory (PNNL) Environmental Molecular Sciences Laboratory (EMSL)
computing group for strong computing support; and the National
Institute of Informatics, and Science Information NETwork 6 (SINET6) for
valuable network support.


\clearpage


\bibliographystyle{JHEP}
\bibliography{./cites/cmb.bib}

\end{document}